\documentclass[12pt]{iopart}

\usepackage{graphicx}
\def\order#1{{\mathcal O}\left(#1\right)}

\begin{document}

\title[Unpolarized QED parton distribution functions in NLO]{Unpolarized QED parton distribution functions in NLO}


\author{A.B. Arbuzov$^{1,2}$, U.E. Voznaya$^{1,2}$}

\address{
$^1$Bogoliubov Laboratory of Theoretical Physics, Joint Institute for Nuclear Research, 
Joliot-Curie, 6, Dubna, 141980, Russia \\
$^2$Dubna State University, Universitetskaya str., 19, Dubna, 141980, Russia
}

\ead{arbuzov@theor.jinr.ru,\ voznaya@theor.jinr.ru} 



\begin{abstract}
Perturbative solutions for unpolarized QED parton distribution and fragmentation functions 
are presented explicitly in the next-to-leading logarithmic approximation.
The scheme of iterative solution of QED evolution equations is described in detail.
Terms up to $\mathcal{O}(\alpha^3L^2)$ are calculated analytically, where
$L=\ln(\mu_F^2/m_e^2)$ is the large logarithm which depends on the factorization energy 
scale $\mu_F\gg m_e$. The results are process independent and relevant 
for future high-precision experiments.
\end{abstract}

\vspace{2pc}



\section{Introduction}\label{sec1}

The studies of high-energy physics processes at future high-luminosity 
electron-positron colliders such as the FCC-ee~\cite{FCC:2018evy} and CEPC~\cite{CEPCStudyGroup:2018ghi} 
require very precise calculations of QED radiative corrections for construction of sufficiently 
accurate theoretical predictions of these processes~\cite{Jadach:2019bye}.
In spite of the existence of powerful methods for calculations in perturbative QED,
it is still very difficult to compute complete radiative corrections in $\mathcal{O}(\alpha^3)$.
On the other hand, we know that higher-order radiative corrections enhanced by the so-called large logarithms 
typically provide the bulk of the effect. For sufficiently inclusive observables, all terms enhanced by
the large logs can be calculated with the help of QED parton distribution function (PDF) approach
which is based on the factorization theorem.

The parton distribution function approach in QED was developed analogously to the QCD one by E.A.~Kuraev and 
V.S.~Fadin~\cite{Kuraev:1985hb} in 1980-s as a step in preparation for high-precision measurements at LEP. 
Actually even earlier, it has been already applied to quasi-elastic neutrino scattering in Ref.~\cite{DeRujula:1979grv}.
The approach is widely used for calculation of QED radiative corrections in high energy physics.
QED evolution equation are just a reduction of the QCD DGLAP (Dokshitzer-Gribov-Lipatov-Altarelli-Parisi)
evolution equations~\cite{Gribov:1972ri,Altarelli:1977zs,Dokshitzer:1977sg} to the abelian case of QED. 
The equations are based on the renormalization group and scale invariance. They allow to effectively account 
the logarithmic dependence on the factorization scale. In QED calculations the corresponding method 
is usually called as the {\em structure function} approach, but here we will adopt the QCD-like notation
in order to preserve the direct correspondence.

The possibility to exponentiate a part of the QED parton distribution functions was
already considered in~\cite{Kuraev:1985hb}, see more details in~\cite{Cacciari:1992pz}. 
The exponentiation is natural because of the 
Yennie-Frautschi-Suura theorem~\cite{Yennie:1961ad}. And it can be done based on the known 
exact solution of the evolution equations in the limit of soft radiation~\cite{Gribov:1972ri}.
But in the present paper we will consider only perturbative order-by-order representations
of QED PDFs, while exponentiation of our results will be presented elsewhere. 

Perturbative solutions of QED evolution equations in the leading logarithmic approximation
are known up to the fifth order for non-singlet electron distribution 
functions~\cite{Przybycien:1992qe,Arbuzov:1999cq}. Singlet higher-order leading log contributions 
to this function were computed in~\cite{Blumlein:2004bs}. One can see that the existing results
for electron PDFs in the collinear leading logarithmic approximation are well cross-checked and
used for estimates of radiative corrections to a wide class of processes. 

As concerns QED PDFs in the next-to-leading logarithmic approximation, it is much less elaborated
and exploited. In the $\order{\alpha^2L}$ spacelike electron PDFs were first considered in 
Ref.~\cite{Berends:1987ab}. Later also timelike electron PDFs in the same order were used to describe
radiative corrections to muon decay spectrum~\cite{Arbuzov:2002cn} and deep inelastic 
scattering~\cite{Blumlein:2002fy}. The resummation in the leading order in the unpolarized case was 
considered in the work \cite{Blumlein:2007kx}. Relatively recently in the series of 
papers~\cite{Blumlein:2011mi,Ablinger:2020qvo,Blumlein:2021jdl}, 
the results for the leading and next-to-leading radiative corrections to electron-positron annihilation
into a virtual photon or $Z$ boson up to the $\order{\alpha^6L^5}$ order were presented.
Those results were obtained with the help of spacelike QED PDFs, but explicit expressions for the 
functions were not given. Moreover, those results are somewhat incomplete since the transitions
from electrons into positrons were missed.

In some recent works the evolution was considered with alternative factorization (Delta, which is DIS-like) and
renormalization ($\alpha(m_Z)$ and $G_{\mu}$) schemes \cite{Frixione:2019lga, Bertone:2019hks, Frixione:2012wtz, Bertone:2022ktl}.
Electrons, and then all "QED flavors", i.e., other charged leptons and quarks were considered.

Leading order QED contributions to the nucleon PDFs have been considered in ref.~\cite{Spiesberger:1994dm} and then applied to nucleon PDFs in refs.~\cite{Roth:2004ti,Martin:2004dh,Ball:2013hta,Bertone:2013vaa}. The QED effects were incorporated into the system of evolution equations for parton distributions of quarks, antiquarks, gluons and photons in a nucleon. Here we will consider only the pure QED case, i.e., QED parton distributions inside electrons, positrons, and photons without inclusion of QCD effects.

In this article,  we describe in detail the iterative solution of evolution equations of parton distribution functions
in QED. The results up to the third iteration are shown explicitly. We consider process-independent PDFs 
which describe the probability density
of finding massless partons (electron, positron or photon) inside electrons and photons. 
Both spacelike and timelike QED PDFs are evaluated.

\section{Master Formula}\label{sec2}

The cross-section of a high-energy $ab\to cd$ process with charged particles in the initial and final states in 
the next-to-leading order (NLO) approximation in pure QED can be represented in the following form~\cite{Arbuzov:2006mu}
\begin{eqnarray} \label{master}
	d \sigma^{\mathrm{NLO}}_{ab\to cd} &=& \sum \limits_{i,j,k,l} \int \limits^{1}_{\bar{z_1}} d z_1 \int \limits^{1}_{\bar{z_2}} 
	d z_2 D^{\mathrm{str}}_{ia} (z_1,\frac{\mu_F^2}{\mu_R^2}) D^{\mathrm{str}}_{jb} (z_2,\frac{\mu_F^2}{\mu_R^2})
	\nonumber \\
	&\times& 
	\left( d \sigma^{(0)}_{ij\to kl} (z_1,z_2) + d \bar{\sigma}^{(1)}_{ij\to kl} (z_1, z_2) 
	+ \mathcal{O}(\alpha^2 L^0)  \right) 
	\nonumber \\
	&\times& \int\limits^{1}_{\bar{y_1}} \frac{d y_1}{Y_1} \int\limits^{1}_{\bar{y_2}} \frac{d y_2}{Y_2}
	D^{\mathrm{frg}}_{ck} (\frac{y_1}{Y_1},\frac{\mu_F^2}{\mu_R^2}) 
	D^{\mathrm{frg}}_{dl} (\frac{y_2}{Y_2},\frac{\mu_F^2}{\mu_R^2}) 
	+ \mathcal{O}\left(\frac{\mu_R^2}{\mu_F^2}\right),
\end{eqnarray}
where $\bar{z_i}$ and $\bar{y_i}$ are some minimal energy fractions defined by experimental conditions; 
$z_i$ are the energy fractions of the incoming partons; $Y_i$ are the energy fractions of the outgoing particles;
$d\sigma^{(0)}_{ij\to kl} (z_1, z_2)$ and $d\bar{\sigma}^{(1)}_{ij\to kl} (z_1, z_2)$ 
are the Born massless parton cross-section and the $\mathcal{O} (\alpha)$ contribution to it.
The bar in $d\bar{\sigma}^{(1)}_{ij\to kl} (z_1, z_2)$ denotes application of a subtraction scheme 
to exclude mass singularities. Here we will use the standard $\overline{\mathrm{MS}}$ scheme. 
The spacelike structure function  $D_{ij}$ (marked "str")  and timelike fragmentation ones (marked "frg") depend on energy 
fractions and the ratio $\mu_F^2/\mu_R^2$, which is the argument 
of the large logarithm.
Note that in QED, the renormalization scale $\mu_R$ is typically chosen to be equal to the electron mass.
The factorization scale $\mu_F$ is usually chosen to be of the order of the energy scale of the hard sub-process.
For example, for the process of electron positron annihilation into a $Z$ boson, we have
$\mu_F=M_Z$ and $\mu_R=m_e$, and the large logarithm is really large numerically, 
$L=\ln(M_Z^2/m_e^2)\approx 24$.
	
As the result, the cross-section in the NLO approximation takes into account
the QED radiative corrections enhanced by the large logarithms and reads
\begin{equation}
	d \sigma^{\mathrm{NLO}}_{ab\to cd} = d\sigma^{(0)}_{ab\to cd} \left\{ 1 + \sum^{\infty}_{k=1} 
	\left(\frac{\alpha}{2\pi}\right)^k \sum^k_{l=k-1} c_{k,l} L^{l}  + \mathcal{O}(\alpha^kL^{k-2}) \right\},
\end{equation}
where $c_{k,l}$ are the coefficients to be computed.
The terms of the type $\alpha^k L^{k}$ provide the leading order (LO) logarithmic approximation, 
and the ones of the type $\alpha^k L^{k - 1}$ yield the NLO contribution.

\section{QED Evolution Equations}\label{sec3}

The main goal of the article is to describe how to compute the
QED parton distribution functions.
Let us consider QED evolution equations for PDFs in the spacelike region. 
The equations are induced by the renormalization group and have the following form, see, 
e.g., Ref.~\cite{Arbuzov:2019hcg}:
\begin{eqnarray} \label{evol_eq}
\!\!\!\!\!\!\!\!\!\!\!\!
 D_{ba}\left(x,\frac{\mu_F^2}{\mu_R^2}\right) = \delta(1-x)\delta_{ba} + \sum\limits_{i=e,\bar{e},\gamma}\int\limits_{\mu_R^2}^{\mu_F^2} 
	\frac{dt \alpha(t)}{2 \pi t}  \int\limits_{x}^{1} \frac{dy}{y} D_{ia}\left(y,\frac{t}{\mu_R^2}\right) P_{bi} \left( \frac{x}{y},t \right), 
\end{eqnarray}
where index $a$ corresponds to the initial particle, e.g., an electron; and 
indices $b$ and $i$ mark QED partons which can be photons $(\gamma)$ 
or massless electrons $(e)$ and positrons $(\bar{e})$.
The parton distribution function $D_{ba} (x,t/\mu_R^2)$ describes the probability  
density to find the massless parton $b$ in the initial (massive) particle $a$
with the of energy fraction $x$ of the initial particle energy at the given energy scale $\sqrt{t}$. 
Note that in the QED PDF formalism 
the initial particle $a$ is a physical on-mass-shell electron, positron or photon, while partons $b$ and $i$ are treated as massless particles of the same types. 
Confusion between, e.g., massive and massless electrons should not appear since their roles are well defined. 
This situation is analogous to the standard QCD PDF formalism, where all partons including even bottom quarks are massless.

$P_{ji}(x,t)$ are called splitting functions or kernels of the evolution equation and describe a perturbative 
transformation of parton $i$ into parton $j$ which takes the energy fraction $x$. 
They can be expanded in a series in the coupling constant $\alpha$
\begin{eqnarray} \label{pee}
	P_{ji} (x,t) = P^{(0)}_{ji} (x) + \frac{\alpha(t)}{2 \pi} P^{(1)}_{ji} (x) + \mathcal{O}(\alpha^2),
\end{eqnarray}
where the running QED coupling constant $\alpha(t)$ is used, see~\ref{App_running_alpha} for details.
    	
The splitting function $P^{(1)}_{ee}(x)$ includes contributions of different sub-processes. 
Expression for this function can be derived from the analogous expression for quark-quark 
function in QCD $P^{(1)}_{qq}(x)$ \cite{Ellis:1996nn} with singlet and non-singlet parts:
\begin{equation}
	P_{q_i q_k}(x,t) = \delta_{ik} P^{NS}_{qq}(x,t) + P^{S}_{qq}(x,t),
\end{equation}
where indices $i$ and $k$ denote quark flavors. In reduction to QED we have only one 
flavor\footnote{The possibility to include in QED, e.g., muons and then have $N_f=2$ is straightforward.}.
The singlet part of splitting functions corresponds to Feynman diagrams with discontinuous 
fermion lines between the initial and final electrons. 
The non-singlet part corresponds to the Feynman diagrams with a continuous electron line. 


The splitting function of electron in positron type $P^{(1)}_{e \bar{e}}(x)$ also have singlet 
and non-singlet parts, so the functions $P^{(1)}_{ee}(x)$ and $P^{(1)}_{e \bar{e}}(x)$ are
\begin{equation}
	P^{(1)}_{ee} = P^{NS}_{ee} + P^{S}_{ee},\qquad
	P^{(1)}_{e \bar{e}} = P^{NS}_{e \bar{e}} + P^{S}_{e \bar{e}}.
\end{equation}
	
There are obvious equalities \cite{Ellis:1996nn} connecting splitting 
functions which involve electrons and positrons because of the $C$ parity 
conservation in QED:
\begin{eqnarray}
	P_{ee} = P_{\bar{e} \bar{e}}, \label{7} \\
	P_{e \bar{e}} = P_{\bar{e} e}, \label{8} \\
	P_{e \gamma} = P_{\bar{e} \gamma}, \label{9} \\
	P_{\gamma e} = P_{\gamma \bar{e}}, \label{10}
\end{eqnarray}
and at the two loop level there is a relation between singlet contributions \cite{Ellis:1996nn}:
\begin{equation}
	P_{e \bar{e}}^S = P_{e e}^S.
\end{equation}
	
We consider spacelike and timelike parton distribution functions (or, in another notation, structure and 
fragmentation functions). Spacelike functions are denoted $\bigl[ P^{(i)} \bigr]_S$, 
and timelike $\bigl[ P^{(i)} \big]_T$. The spacelike functions correspond to transitions 
from massive particles into massless ones, and the timelike ones correspond to transition from massless 
particles into massive ones. The difference between timelike and spacelike functions appear in QED only 
in NLO \cite{Arbuzov:2019hcg} and only in the $\Theta$ parts. In the leading order spacelike and timelike 
functions are equal. The $\Delta$ parts of timelike and spacelike functions are also equal. The $\Theta$ part of a splitting function or a parton sub-process cross section corresponds to the situation of real radiation which leads to an energy loss so that the remaining energy fraction $x<1-\Delta$. The $\Delta$ part incorporates effects due to virtual (loop) corrections and soft radiation so that the remaining energy fraction $x$ is exactly one or very close to this value: $1-\Delta\leq x\leq 1$. Note that the limit $\Delta\to 1$ has to be taken. Here we will do that analytically.
There is the so-called Gribov-Lipatov relation~\cite{Gribov:1972rt} between spacelike and timelike functions:
\begin{equation}
- x \bigl[ P_{ba}^{(0)} \bigr]_T \left(\frac{1}{x} \right) = \bigl[ P_{ab}^{(0)} \bigr]_S (x) + \Delta(x),
\end{equation}
where the $\Delta(x)$ term can be calculated using analytical continuation of the timelike 
function to the nonphysical region $x >1$ \cite{Curci:1980uw}. 
For electron-electron function, the correction to the Gribov-Lipatov relation reads
\begin{eqnarray}
&& \Delta_{ee}(x) = \left( \frac{\alpha}{2 \pi}\right)^2 \biggl[ 4 \frac{1+x^2}{1-x} \ln x\ln (1-x) \nonumber \\
&& +  \left( \frac{6}{1-x} - 5 - 5x\right) \ln x + \left(1 + x - 2 \frac{1+x^2}{1-x} \right) \ln^2 x  \biggr].
\end{eqnarray}

\subsection{Initial conditions}
	
We are going to solve evolution equations~(\ref{evol_eq}) by iterations. 
So we need some initial conditions. In the NLO approximation in the $\overline{\mathrm{MS}}$
scheme, they are \cite{Arbuzov:2002rp, Blumlein:2011mi}
\begin{eqnarray} \label{IC}
	&& D_{ee}^{(0)}(x,\mu_R^2/m_e^2) = \delta (1-x) + \frac{\alpha}{2\pi}d_{ee}^{(1)}(x,\mu_R^2/m_e^2), \\
	&& D_{\gamma e}^{(0)} (x,\mu_R^2/m_e^2) = \frac{\alpha}{2\pi}d_{\gamma e}^{(1)}(x,\mu_R^2/m_e^2), \\
	&& D_{e \gamma}^{(0)} (x,\mu_R^2/m_e^2) = \frac{\alpha}{2\pi}d_{e \gamma}^{(1)}(x,\mu_R^2/m_e^2), \\
	&& D_{e\bar{e}}^{(0)}(x,\mu_R^2/m_e^2)  = \frac{\alpha}{2\pi}d_{e\bar{e}}^{(1)}(x,\mu_R^2/m_e^2) = 0,
\end{eqnarray}
\begin{eqnarray}
	d_{ee}^{(1)}(x,\mu_R^2/m_e^2)  &=& \left[\frac{1 +x^2}{1-x}\biggl(\ln \frac{\mu_R^2}{m_e^2}- 1 - 2 \ln(1-x) \biggr)\right]_+ 
	\nonumber \\
	&=&  \left[\frac{1 +x^2}{1-x}(- 1 - 2 \ln(1-x) )\right]_+,
\end{eqnarray}
note that here and in what follows we apply the natural choice of the QED renormalization constant 
$\mu_R = m_e$ \cite{Blumlein:2011mi}, 
\begin{eqnarray}
	&& d_{\gamma e}^{(1)}(x) = -\frac{1 + (1-x)^2}{ x}(2 \ln x + 1), \\
	&& d_{e \gamma}^{(1)}(x) = 0.
\end{eqnarray}
They are fixed with the help
of the known results of perturbative calculations. Using the complete result
in the  first order $\mathrm{O}(\alpha)$ we can fix functions $d^{(1)}_{ba}(x)$
in a chosen subtraction scheme. In any case, the following equality should be held
\begin{equation} \label{fixing}
	\sigma^{(1)}_{e\bar{e}\to \gamma^*}(x) = \bar{\sigma}^{(1)}_{e\bar{e}\to \gamma^*}(x) 
	+ 2 \frac{\alpha}{2\pi}\left[ P^{(0)}_{ee}(x) L + d_{ee}^{(1)}(x) \right]  \otimes \sigma^{(0)}_{e \bar{e}}(x)
	+ \mathcal{O}\left(\frac{m_e^2}{\mu_F^2}\right),
\end{equation}
where $\sigma^{(1)}$ is the complete one-loop contribution to the electron-positron annihilation process 
into a virtual photon $\bar{e}+e\to \gamma^*$ in the standard QED with massive electrons. 
Here $x$ is the energy fraction of the
produced virtual photon with respect to the total c.m.s. energy of the initial particles.
To restore the mass dependence in the results of calculations within a massless 
theory (up to $\mathcal{O}(m_e^2/\mu_F^2)$ terms) one can use the so-called {\it massification} procedure. 
The above equation is an example of this procedure.
	
We can fix the next-to-leading contributions to the splitting functions 
and the second-order $d^{(2)}_{ba}$ functions
with the help 
of a complete $\mathcal{O}(\alpha^2)$ result, e.g., for the same process 
of electron-positron annihilation with initial
state radiation corrections~\cite{Berends:1987ab, Blumlein:2011mi}. 
Note that separation of terms on the right-hand side of Eq.~(\ref{fixing})
depends on the chosen scheme and factorization scale.

	
\subsection{Iterative solution}
	
Evolution equations can be solved using the iterative method
with the initial conditions~(\ref{IC}).
On the first step, we substitute the initial conditions for $D_{ia}(e,s')$ in 
the evolution equations~(\ref{evol_eq}) under the integral on the right-hand side.
The result of the first iteration reads\footnote{We dropped arguments of functions on the right hand side.}
\begin{eqnarray}
	&& D_{ee}^{\mathrm{(I)}}(x,\mu_F^2/\mu_R^2) = D_{ee}^{(0)} + \frac{\alpha}{2 \pi} \left( P_{ee} \otimes D_{ee}^{(0)} 
	+ P_{e \gamma} \otimes D_{\gamma e}^{(0)} + P_{e \bar{e}} \otimes D_{\bar{e} e}^{(0)} \right) 
	\nonumber \\
	&& \quad = \delta (1-x) + \frac{\alpha}{2 \pi}  d_{ee}^{(1)} 
	+ \frac{\alpha}{2 \pi}  \left( P_{ee}^{(0)} + \frac{\alpha}{2 \pi} P_{ee}^{(1)} \right) 
	\otimes \left( \delta (1-x) + \frac{\alpha}{2 \pi}  d_{ee}^{(1)} \right)
	\nonumber \\
	&& \quad  + \biggl(P_{e \gamma}^{(0)} + \frac{\alpha}{2 \pi} P_{e \gamma}^{(1)} \biggr) \otimes d_{\gamma e}^{(1)} = \delta (1-x)  + \frac{\alpha}{2 \pi} d_{ee}^{(1)} + \frac{\alpha}{2 \pi}  L  P_{ee}^{(0)} , 
	\\ 
	&& D_{\gamma e}^{\mathrm{(I)}}(x,\mu_F^2/\mu_R^2) = D_{\gamma e}^{(0)} + \frac{\alpha}{2 \pi} (P_{\gamma \gamma} 
	\otimes D_{\gamma e}^{(0)}  + P_{\gamma e} \otimes D_{ee}^{(0)} + P_{\gamma \bar{e}} \otimes D_{\bar{e} e}^{(0)})
	\nonumber \\ 
	&& \quad = \frac{\alpha}{2 \pi} d_{\gamma e}^{(1)} + \frac{\alpha}{2 \pi}  \left( P_{\gamma \gamma}^{(0)}  + \frac{\alpha}{2 \pi} P_{\gamma \gamma}^{(1)} \right) \otimes \frac{\alpha}{2 \pi} d_{\gamma e}^{(1)} 	\nonumber \\ 
	&& \quad
	+ \frac{\alpha}{2 \pi}  \left( P_{\gamma e}^{(0)} 
       + \frac{\alpha}{2 \pi} P_{\gamma e}^{(1)} \right) \otimes \bigl(\delta (1-x)
     + \frac{\alpha}{2 \pi} d_{ee}^{(1)}\bigr) \nonumber \\ 
	&& \quad= \frac{\alpha}{2 \pi} d_{\gamma e}^{(1)} +  \frac{\alpha}{2 \pi} L  P_{\gamma e}^{(0)},
	\\
	&& D_{e \bar{e}}^{\mathrm{(I)}} (x,\mu_F^2/\mu_R^2) = D_{e \bar{e}}^{(0)} + \frac{\alpha}{2 \pi} (P_{ee}^{(0)} \otimes D_{e \bar{e}}^{(0)}+ P_{e \bar{e}}^{(0)} \otimes D_{ee}^{(0)} + P_{e \gamma}^{(0)} \otimes D_{e \gamma}^{(0)}) \nonumber \\ 
	&& \quad = \frac{\alpha}{2 \pi} P_{e \bar{e}}^{(0)} = 0. 
\end{eqnarray}

On the second iteration step, we put the result of the first step under the integral~(\ref{evol_eq}). 
The results of the second and the third iterations are presented in~\ref{secB1}.

We also show the results of iterations as functions of $z$. The difference between spacelike and timelike 
functions is only in $P_{ij}^{(1)}$, so the parton distributions $D_{ij}^{(1)}$ are equal for spacelike 
and timelike cases. As we are interested only in LO an NLO contributions (in other words, LL and NLL approximation), we can omit terms proportional 
to $\left( \frac{\alpha}{2 \pi} \right)^{k+2} L^k$. The results are presented in~\ref{secC1}.
We used splitting functions and massless Wilson coefficients~\cite{Blumlein:2011mi,Furmanski:1980cm,Ellis:1996nn,Berends:1987ab}
which are given in~\ref{secA1}.

\subsection{Numerical results}
\begin{figure}[ht]
		\includegraphics[width=0.9\linewidth]{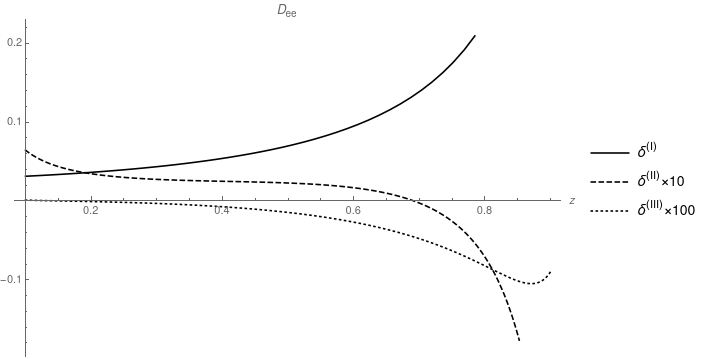}
		\caption{Difference between the results of three iterations for the function $\bigl[ D_{ee} (z,\mu) \bigr]_S$ for $0.1<z<0.9$.} \label{3it}
\end{figure}

We can illustrate the convergence of the constructed iterative solution for the {\em electron in electron} spacelike PDF by looking at the differences of three iterations:
which are shown in Fig.~\ref{3it} for $L = 24$ $(\mu_F\approx M_Z)$ and $\alpha = 1/137$.
Note that by looking at this plot one can only roughly estimate the size of the corresponding
radiative corrections, since convolution with a kernel cross section should follow. The grows of the magnitude of corrections at $z\to1$ comes from the infrared singularity which is removed after adding the contributions
of soft and virtual emission, which are given by the terms proportional to $\delta(1-z)$.

\section{Conclusions}
	
In this way, we described the iterative method of solving QED evolution equations. Extension of the presented
calculations to higher orders is straightforward. 
Note that our results are scheme and factorization scale dependent. Application of other schemes and variation
of renormalization and factorization scales are possible within the same line of calculations.
Here, we presented only the electron parton distribution functions, photon PDFs will be presented
elsewhere.

The calculated QED parton distribution functions can be applied to various high-energy
processes with participation of electrons and/or positrons. Note that the PDFs here
are treated in the collinear approximation which should match the experimental conditions
of particle registration of the process under study. In other words, the physical observable
should not be sensitive to transverse momenta of initial and/or final state radiation.
In any case, the results obtained within the collinear approximation will serve as benchmarks 
for Monte Carlo simulations which describe the complete kinematics of events.

As we mentioned in the introduction, our results partially intersect with the ones
given in~Refs.~\cite{Blumlein:2011mi,Ablinger:2020qvo,Blumlein:2021jdl}. 
Results a detailed comparison will be presented later.

\subsection{Acknowledgments}
A.A. is grateful for support 
to the Russian Science Foundation, project No. 22-12-00021.

\appendix
\setcounter{section}{0}

\section{Running coupling constant}
\label{App_running_alpha}
	
The expression for the QED running coupling constant at a certain energy scale $\mu$ in the 
$\overline{\mathrm{MS}}$ scheme can be found, e.g., in~\cite{Baikov:2012rr,Gorishnii:1991hw},
\begin{equation}
\alpha(\mu^2) = \frac{{\alpha(\mu_R^2)}}{1 + \overline{\Pi}(\mu,\mu_R,\alpha(\mu_R^2))}\, ,
\end{equation}
with 
\begin{eqnarray} \label{polar}
	\overline{\Pi}\left(\mu,\mu_R,\alpha(\mu_R^2)\right) = \frac{\alpha(\mu_R^2)}{\pi} \left( \frac{5}{9}-\frac{L}{3} \right) 
	+ \left(\frac{\alpha(\mu_R^2)}{\pi}\right)^2 \left( \frac{55}{48}-\zeta_3-\frac{L}{4} \right) \nonumber \\
	+ \left(\frac{\alpha(\mu_R^2)}{\pi}\right)^3 \left( -\frac{L^2}{24} \right) + \ldots
\end{eqnarray}
where $L =\ln(\mu^2/\mu_R^2)$ is again a large logarithm.
After expansion, we get
\begin{eqnarray}
	\alpha(\mu^2) &=&  \alpha (0) \biggl\{ 1 +\frac{\alpha (0)}{2 \pi}  
	\left(  - \frac{10}{9} + \frac{2}{3} L \right) 
	+ \left( \frac{\alpha (0)}{2 \pi} \right)^2  \biggl(  - \frac{1085}{324} + 4 \zeta_3  \nonumber \\ 
	&-& \frac{13}{27} L
	+ \frac{4}{9} L^2 \biggr)  + \mathcal{O} \left( \alpha^3(0)\right) \biggr\},
\end{eqnarray}
where $\zeta_n\equiv\zeta(n)$ is the Riemann zeta function.
Here we put $\mu_R=m_e$ and assume $\alpha(m_e^2)\approx\alpha(0)\equiv\alpha$.
To avoid double counting and following the QCD formalism, we will omit the $-10/9$ 
non-logarithmic term in the running of $\alpha$ since the corresponding effect 
is already taken into account in splitting functions $P^{(1)}_{ji}$.

\section{Convolution and plus prescription}
\label{App_conv}

The evolution equations involve the convolution operation 
\begin{equation} \label{convolution}
	\bigl(f\otimes g\bigr)(x) \equiv \int \limits^1_0 dz \int \limits^1_0 dy f(z) g(y) \delta(x-yz)
	= \int \limits^1_x \frac{d z}{z} f(z) g(\frac{x}{z}).
\end{equation}
Many of relevant functions have a regularization at $x\to 1$ and can be represented as
a sum of the so-called $\Theta$ and $\Delta$ parts:
\begin{eqnarray} \label{DeltaTheta}
	f(x) = \lim_{\Delta\to 0}\biggl(f_\Theta(x)\Theta(1-x-\Delta)+f_\Delta\delta(1-x)\biggr),
\end{eqnarray}
where $\Theta(x)$ is the standard Heaviside step function.
In many cases, such functions can be regularized 
with the help of the {\it plus prescription} which acts in
an integral with a regular function as follows
\begin{equation} \label{PlusPres}
	\int \limits_{z}^{1} d x [f(x)]_+ g(x) = \int \limits_{0}^{1} d x f(x)\biggl[g(x)\Theta(x-z) - g(1)\biggr].
\end{equation}
If function $f(x)$ is regularized by the plus prescription, it satisfies the sum rule
\begin{equation} \label{sum_rule}
	f_{\Delta} = -\int \limits^{1-\Delta}_0 f_{\Theta}(z) d z.
\end{equation}
For the convolution of two functions with $\Delta$ parts we get
\begin{equation}
	\Bigl( f \otimes g \Bigr)_{\Theta} (z) = \lim_{\Delta \to 0} 
	\Bigl\{ \int \limits_{z/{(1 - \Delta)}}^{1-\Delta} \frac{d x}{x} f_{\Theta} (x)
	g_{\Theta} \left(\frac{z}{x} \right)  + f_\Delta g_{\Theta} (z) + f_{\Theta} (z) g_\Delta \Bigr\}.
\end{equation}
Note that in this way we get only the $\Theta$ part of the convolution.
If the result satisfies sum rule~(\ref{sum_rule}), its $\Delta$ part
is easily restored. 
If the sum rule is not applicable, the $\Delta$ part of a convolution  
can be calculated as
\begin{equation}
	\left( f \otimes g \right)_\Delta = f_\Delta  g_\Delta 
	- \int \limits^{1}_{1-\Delta} dy f(y) \int \limits^{\frac{1-\Delta}{y}}_{1-\Delta} g(x) dx,
\end{equation}
as follows from the definitions~(\ref{convolution}) and (\ref{DeltaTheta}).

\section{Splitting functions}
\label{secA1}

Process-independent parton splitting functions in the lowest order are
	\begin{eqnarray} 
	&& P_{ee}^{(0)} (x) = \left[\frac{1 + x^2}{1-x}\right]_+, \qquad
	P_{e\gamma}^{(0)} (x) = x^2 + (1- x)^2, \qquad
	P_{e \bar{e}}^{(0)} (x) = 0, \nonumber \\
	&& P_{\gamma e}^{(0)} (x) = \frac{1 + (1-x)^2}{x}, \qquad
	P_{\gamma\gamma}^{(0)} (x) = \frac{\beta_0}{2} \delta (1-x). \\ \nonumber
	\end{eqnarray}
Spacelike functions read
         \begin{eqnarray}
	&& \bigl[P_{ee}^{(1,NS)}\bigr]_S (x) = C_f^2 \left( \left(-2 \ln x \ln (1-x) - \frac{3}{2} \ln x \right) \frac{1+x^2}{1-x} - \left( \frac{3}{2}+ \frac{7}{2} x \right) \ln x 
	\right. \nonumber \\ \nonumber && \quad\left. 
	- \frac{1}{2}(1+x) \ln^2 x - 5 (1-x) \right)
	+ C_f T_f \left( - \frac{2}{3} \ln x- \frac{10}{9} \frac{1+x^2}{1-x} - \frac{4}{3} (1-x) \right)   \\
	&& \quad+\delta(1-x) \left( C_f^2 \left( \frac{3}{8} - \frac{\pi}{2} +6 \zeta_3 \right) - C_f T_f \left( \frac{1}{6} +\frac{2 \pi}{9} \right) \right), 
	\\ \nonumber
	&& \bigl[P_{ee}^{(1,S)}(x)\bigr]_S = C_f T_f \Biggl( \frac{20}{9 x} - 2 + 6 x - \frac{56}{9} x^2 + (1 + 5 x+\frac{8}{3} x^2) \ln x 	\\ 
	&&  \quad - (1+x)\ln^2 x \Biggr), 
	\\ \nonumber
 	&& P_{e \bar{e}}^{(1, NS)}(x) = C_f^2 \left( 2 \frac{1+x^2}{1+x} S_2 (x) + 2 (1+x) \ln x +4(1-x) \right),
	\\ 
	&& \bigl[ P_{ee}^{(1)}(x)\bigr]_S = \bigl[P_{ee}^{(1,NS)}\bigr]_S (x) + \bigl[P_{ee}^{(1,S)}\bigr]_S (x),
 	\\ 
	&& \bigl[ P_{e \bar{e}}^{(1)}(x)\bigr]_S = P_{e \bar{e}}^{(1,NS)} (x) + \bigl[P_{ee}^{(1,S)}\bigr]_S (x),
	\\ 
	&& \bigl[P_{e\gamma}^{(1)}\bigr]_S (x) = C_f T_f \biggl[ 4 - 9 x - (1-4 x) \ln x - (1-2x) \ln^2 x 
	 \nonumber \\ &&\quad  
	+ 4 \ln (1-x)  
	+ \biggl( 2 ( \ln (1-x)- \ln x)^2 - 4 ( \ln (1-x)- \ln x ) 
	 \nonumber \\ &&\quad  
        - 4 \zeta (2) + 10 \biggr) P^{(0)}_{e\gamma}(x) \biggr], 
	\\ \nonumber
	&& \bigl[P_{\gamma e}^{(1)}\bigr]_S (x) = C_f^2 \biggl( - \frac{5}{2} - \frac{7}{2} x + \left( 2+ \frac{7}{2} x \right) \ln x 
	- \left( 1 - \frac{1}{2} x \right) \ln x^2   \\  &&\quad 
	- 2 x \ln (1-x) - ( 3 \ln (1-x) + \ln^2 (1-x) )  P^{(0)}_{\gamma e}(x) \biggr)  \nonumber
 	\\ 
	&& \quad
	+ C_f \Biggl(- \frac{4x}{3} - \left( \frac{20}{9} + \frac{4\ln (1-x)}{3}  \right) P^{(0)}_{\gamma e}(x) \Biggr). 
	\end{eqnarray}
The timelike functions read
        \begin{eqnarray} 
	&& \bigl[ P_{ee}^{(1, NS)}(x) \bigr]_T= C_f^2 \left( \left(2 \ln x \ln (1-x) + \frac{3}{2} \ln x  - 2\ln^2 x\right) \frac{1+x^2}{1-x} - \left( \frac{7}{2}+ \frac{3}{2} x \right) \ln x 
	\right. \nonumber \\ \nonumber && \quad\left. 
	- \frac{1}{2}(1+x) \ln^2 x - 5 + 3x \right)
	+ C_f T_f \left( \left(- \frac{2}{3} \ln x- \frac{10}{9} \right) \frac{1+x^2}{1-x} - \frac{4}{3} (1-x) \right)   \\
	&& \quad+\delta(1-x) \left( C_f^2 \left( \frac{3}{8} - \frac{\pi}{2} +6 \zeta_3 \right) - C_f T_f \left( \frac{1}{6} +\frac{2 \pi}{9} \right) \right), 
	\\ \nonumber
        && \bigl[ P_{ee}^{(1, S)}(x) \bigr]_T = C_f T_f \Biggl(\ln x \left(  - 5 - 9 x - \frac{8}{3} x^2 \right) + \ln^2 x ( 1 + x )
  - 8 - \frac{20}{9 x} 	\\ 
        && \quad  + 4x + \frac{56}{9} x^2\Biggr) , \\ \nonumber
        && \bigl[ P_{ee}^{(1)}(x)\bigr]_T = \bigl[P_{ee}^{(1,NS)}\bigr]_T (x) + \bigl[P_{ee}^{(1,S)}\bigr]_T (x),
        \\ \nonumber
	&& \bigl[ P_{e \gamma}^{(1)}(x) \bigr]_T= T_f^2 \left[ - \frac{8}{3} - \left(\frac{16}{9} + \frac{8}{3} \ln x + \frac{8}{3} \ln (1-x) \right)\right] +
 C_f N_f \left[ -2 + 3x      \right.   \\ 
	&& \quad  + (-7 + 8x) \ln x - 4 \ln (1-x) + (1 - 2 x) \ln^2 x + \left( -4 \ln x\ln (1-x) - 2 \ln^2 x   \right.  \nonumber \\ 
	&& \quad \left.  - 2 \ln (1-x) + 2 \ln x - 2 \ln^2 (1-x) + 16 S_1 (x) + 12 \zeta_2 - 10 \right) P_{e \gamma}^{(0)}\left. \right],
        \\ \nonumber
	&& \bigl[ P_{\gamma e}^{(1)}(x) \bigr]_T= C_f^2 \left[ -\frac{1}{2} + \frac{9}{2} x \left( -8 + \frac{1}{2} x \right) \ln x + 2 x \ln (1-x) + \left( 1 - \frac{1}{2}\right) \ln^2 x  \right. \\
	&& \quad \left. + \left( \ln^2 (1-x) + 4 \ln x \ln (1-x) - 8 S_1 (x) - \frac{4}{3} \pi \right) P_{\gamma e}^{(0)}\right],
	\\ 
	&& S_2 (x) = \int \limits^{\frac{1}{1+x}}_{\frac{x}{1+x}} \frac{d z}{z} \ln \frac{1-z}{z} = - 3 \zeta_2 - \frac{\ln^2 x}{2} + 2 \ln x \ln (1+x) + 2 \mathrm{Li}_2 (1+x);
	\\ 
	&& S_1 (x) = \int \limits^{1-x}_{0} \frac{d z}{z} \ln (1-z),
	\end{eqnarray}
	where $P_{ee}^{NS}(x)$ and $P_{ee}^{S}(x)$ are non-singlet and singlet contributions, which are different for spacelike and timelike functions. $P_{e \bar{e}}^{(1, NS)}(x) $ 
	is the same in the spacelike and timelike cases.
	Note the in QED the QCD constants are reduced to $C_f=1$ and $T_f = 1$; $\beta_0 = -\frac{4}{3}$.

\section{PDFs in terms of convolutions}
\label{secB1}

Here we present results for electron PDFs in terms of convolutions.

\begin{eqnarray}
	&& D_{ee}^{(\mathrm{II})}(x,\mu_F^2/\mu_R^2) = D_{ee}^{(\mathrm{I})}
	+ \left(\frac{\alpha}{2 \pi} \right)^2 L \biggl( d_{\gamma e}^{(1)}(x) \otimes P_{e \gamma}^{(0)}
	 + P_{ee}^{(1)} 
        \nonumber \\ && \quad  -  \frac{10}{9} P_{ee}^{(0)} 	+ P_{ee}^{(0)} \otimes d_{ee}^{(1)}(x) \biggr) 
 	\nonumber \\ && \quad 
       + \Bigl( \frac{\alpha}{2 \pi} \Bigr)^2 L^2 
	\Bigl( \frac{1}{2} P_{\bar{e} e}^{(0)} \otimes P_{e \bar{e}}^{(0)} + \frac{1}{2} P_{\gamma e}^{(0)} \otimes P_{e \gamma}^{(0)} + \frac{1}{3} P_{ee}^{(0)} +
	  \frac{1}{2} P_{ee}^{(0)} \otimes P_{ee}^{(0)} \Bigr),
	\\  
	&& D_{\gamma e}^{(\mathrm{II})} (x,\mu_F^2/\mu_R^2) = D_{\gamma e}^{(\mathrm{I})} 
	+ \Bigl( \frac{\alpha}{2 \pi} \Bigr)^2 L  \Bigl( P_{\gamma e}^{(1)} 
	+ d_{\gamma e}^{(0)} \otimes P_{\gamma \gamma}^{(0)} 
    - \frac{10}{9} P_{\gamma e}^{(0)}  	+ d_{ee}^{(1)}(x) \otimes P_{\gamma e}^{(0)} \Bigr) 
    \nonumber \\ && \quad 
    + \left(\frac{\alpha}{2 \pi} \right)^2 L^2 
	\biggl( \frac{1}{3} P_{\gamma e}^{(0)} + \frac{1}{2} P_{\gamma e}^{(0)} \otimes P_{\gamma \gamma}^{(0)} +
	  \frac{1}{2} P_{ee}^{(0)} \otimes P_{\gamma e}^{(0)} + \frac{1}{2} P_{\bar{e} e}^{(0)} \otimes P_{\gamma \bar{e}}^{(0)}\biggr),
    \\  
    && D_{e \bar{e}}^{(\mathrm{II})} (x,\mu_F^2/\mu_R^2) = D_{e \bar{e}}^{(\mathrm{I})}  + \left( \frac{\alpha}{2 \pi}\right)^2 L \left( P_{e \bar{e}}^{(1)} - \frac{10}{9} P_{e \bar{e}}^{(0)} + d_{\gamma e}^{(1)} \otimes P_{e \gamma}^{(0)} + d_{ee}^{(1)} \otimes P_{e \bar{e}}^{(0)} \right) 
    \nonumber \\ && \quad 
    + \left( \frac{\alpha}{2 \pi}\right)^2 L^2 \left( \frac{1}{3} P_{e \bar{e}}^{(0)} + \frac{1}{2} P_{e \gamma}^{(0)} \otimes P_{\gamma \bar{e}}^{(0)} + \frac{1}{2} P_{ee}^{(0)} \otimes P_{e \bar{e}}^{(0)} + \frac{1}{2} P_{\bar{e} \bar{e}}^{(0)} \otimes P_{e \bar{e}}^{(0)}  \right).
	\end{eqnarray}

	\begin{eqnarray}
	&& D_{ee}^{(\mathrm{III})} (x,\mu_F^2/\mu_R^2) = D_{ee}^{(\mathrm{II})} + \left( \frac{\alpha}{2 \pi} \right)^3 L^2   \Bigl( \frac{1}{2}  P_{\gamma e}^{(1)} \otimes P_{e \gamma}^{(0)} + \frac{1}{2} P_{\bar{e} e}^{(0)} \otimes P_{e \bar{e}}^{(1)} + \frac{1}{3}  d_{\gamma e}^{(1)} \otimes P_{e \gamma}^{(0)} 
	\nonumber \\ && \quad 
	+ \frac{1}{2}  d_{\gamma e}^{(1)} \otimes P_{\gamma \gamma}^{(0)} \otimes P_{e \gamma}^{(0)} + \frac{1}{2}  P_{\gamma e}^{(0)} \otimes P_{e \gamma}^{(1)} - \frac{10}{9}  P_{\gamma e}^{(0)} \otimes  P_{e \gamma}^{(0)} + \frac{2}{3}  P_{ee}^{(1)}
	\nonumber \\
	&& \quad 
  + \frac{1}{2} P_{\bar{e} e}^{(1)} \otimes P_{e \bar{e}}^{(0)} - \frac{10}{9} P_{\bar{e} e}^{(0)} \otimes P_{e \bar{e}}^{(0)} + \frac{1}{2} d_{\gamma e}^{(1)} \otimes P_{\bar{e} \gamma}^{(0)} \otimes P_{e \bar{e}}^{(0)} + \frac{1}{2} d_{ee}^{(1)} \otimes P_{\bar{e} e}^{(0)} \otimes P_{e \bar{e}}^{(0)}
	\nonumber \\
	&& \quad 
	 + \frac{1}{2}  d_{ee}^{(1)} \otimes P_{\gamma e}^{(0)} \otimes P_{e \gamma}^{(0)} - \frac{13}{54}   P_{ee}^{(0)} + \frac{1}{2}  P_{ee}^{(0)} \otimes d_{\gamma e}^{(1)} \otimes P_{e \gamma}^{(0)} + P_{ee}^{(0)} \otimes P_{ee}^{(1)}  
	\nonumber  \\
	&& \quad
	+ \frac{1}{3}  P_{ee}^{(0)} \otimes d_{ee}^{(1)} - \frac{10}{9}  P_{ee}^{(0)} \otimes P_{ee}^{(0)} + \frac{1}{2}  P_{ee}^{(0)} \otimes P_{ee}^{(0)} \otimes d_{ee}^{(1)} \Bigr) 
	\nonumber \\
	&& \quad
 +\left( \frac{\alpha}{2 \pi} \right)^3 L^3   \left( \frac{1}{3}  P_{\bar{e} e}^{(0)} \otimes P_{e \bar{e}}^{(0)} + \frac{1}{6}  P_{\bar{e} \bar{e}}^{(0)} \otimes P_{e \bar{e}}^{(0)} \otimes  P_{e \bar{e}}^{(0)}+ \frac{4}{27}  P_{ee}^{(0)} + \frac{1}{6} P_{\gamma e}^{(0)} \otimes P_{e \bar{e} }^{(0)} \otimes P_{\bar{e} \gamma}^{(0)} \right.
 	\nonumber \\
	&& \quad
	\left.+ \frac{1}{3}  P_{ee}^{(0)} \otimes  P_{\bar{e} e}^{(0)} \otimes P_{e \bar{e}}^{(0)} + \frac{1}{6} P_{\gamma \bar{e}}^{(0)} \otimes P_{\bar{e} e}^{(0)} \otimes P_{e \gamma}^{(0)} \right.
	 +\frac{1}{3}  P_{\gamma e}^{(0)} \otimes  P_{e \gamma}^{(0)} + \frac{1}{6}  P_{\gamma e}^{(0)} \otimes P_{\gamma \gamma}^{(0)} \otimes P_{e \gamma}^{(0)}     	\nonumber \\ && \quad \left. 
	+ \frac{1}{3}  P_{ee}^{(0)} \otimes  P_{\gamma e}^{(0)} \otimes P_{e \gamma}^{(0)}
	+\frac{1}{3}  P_{ee}^{(0)} \otimes P_{ee}^{(0)} + \frac{1}{6}  P_{ee}^{(0)} \otimes P_{ee}^{(0)} \otimes P_{ee}^{(0)} \right),
	\end{eqnarray}
	
	\begin{eqnarray}
	&& D_{\gamma e}^{(\mathrm{III})} (x,\mu_F^2/\mu_R^2) = D_{\gamma e}^{(\mathrm{II})}   + \left( \frac{\alpha}{2 \pi}  \right)^3 L^2   \left(   \frac{2}{3} P_{ \gamma e}^{(1)}+ \frac{1}{2} P_{ \gamma e}^{(1)} \otimes P_{ \gamma  \gamma }^{(0)} + \frac{1}{2} P_{ \gamma \bar{e} }^{(0)} \otimes P_{\bar{e} e}^{(1)}\right.
	\nonumber \\ && \quad
   - \frac{10}{9} P_{ \gamma \bar{e} }^{(0)} \otimes P_{\bar{e} e}^{(0)} 	+ \frac{1}{2} P_{ \gamma \bar{e}}^{(1)} \otimes P_{\bar{e} e}^{(0)} + \frac{1}{3} d_{ \gamma e}^{(1)} \otimes P_{ \gamma  \gamma }^{(0)} + \frac{1}{2} d_{ \gamma e}^{(1)} \otimes \left(P_{ \gamma  \gamma }^{(0)}\right)^2 
    \nonumber \\ && \quad
+ \frac{1}{2} d_{ \gamma e}^{(1)} \otimes P_{ \gamma \bar{e} }^{(0)} \otimes P_{\bar{e}  \gamma }^{(0)} - \frac{13}{54} P_{ \gamma e}^{(0)}  + \frac{1}{2} P_{ \gamma e}^{(0)} \otimes P_{ \gamma  \gamma }^{(1)} - \frac{10}{9} P_{ \gamma e}^{(0)} \otimes P_{ \gamma  \gamma }^{(0)} 
	\nonumber \\ && \quad
+ \frac{1}{2} P_{ \gamma e}^{(0)} \otimes d_{\gamma e}^{(1)} \otimes P_{e \gamma }^{(0)} + \frac{1}{2} P_{ee}^{(1)} \otimes P_{ \gamma e}^{(0)}  + \frac{1}{2} d_{ee}^{(1)} \otimes P_{ \gamma \bar{e} }^{(0)} \otimes P_{\bar{e} e}^{(0)}
\nonumber \\ && \quad
+ \frac{1}{3} d_{ee}^{(1)} \otimes P_{ \gamma e}^{(0)} + \frac{1}{2} d_{ee}^{(1)} \otimes P_{ \gamma e}^{(0)} \otimes P_{ \gamma  \gamma }^{(0)} + \frac{1}{2} P_{ee}^{(0)} \otimes P_{ \gamma e}^{(1)}
\nonumber \\ && \quad
- \frac{10}{9} P_{ee}^{(0)} \otimes P_{ \gamma e}^{(0)} \left. + \frac{1}{2} P_{ee}^{(0)} \otimes d_{ee}^{(1)} \otimes P_{ \gamma e}^{(0)} \right)
\nonumber \\ && \quad
+\left( \frac{\alpha}{2 \pi}  \right)^3 L^3   \left(  \frac{1}{3} P_{ \gamma \bar{e} }^{(0)} \otimes P_{ \bar{e} e}^{(0)}  \right.
+ \frac{1}{6} P_{ \gamma \bar{e} }^{(0)} \otimes P_{\bar{e} \bar{e}}^{(0)} \otimes P_{e \bar{e} }^{(0)}  + \frac{1}{6} P_{ \gamma e }^{(0)} \otimes P_{\bar{e} e}^{(0)} \otimes P_{e \bar{e} }^{(0)} 
\nonumber \\ && \quad
+ \frac{1}{6} P_{ \gamma \bar{e} }^{(0)} \otimes P_{\bar{e} e}^{(0)} \otimes P_{ \gamma  \gamma }^{(0)} + \frac{4}{27} P_{ \gamma e}^{(0)} + \frac{1}{3} P_{ \gamma e}^{(0)} \otimes P_{ \gamma  \gamma }^{(0)}  + \frac{1}{6} P_{ \gamma e}^{(0)} \otimes P_{ \gamma  \gamma }^{(0)} \otimes P_{ \gamma  \gamma }^{(0)}
\nonumber \\ && \quad
 + \frac{1}{6} P_{ \gamma e}^{(0)} \otimes P_{ \gamma \bar{e} }^{(0)} \otimes P_{\bar{e}  \gamma }^{(0)} + \frac{1}{6} P_{ \gamma e}^{(0)} \otimes P_{ \gamma e}^{(0)} \otimes P_{e \gamma }^{(0)}  + \frac{1}{6} P_{ee}^{(0)} \otimes P_{ \gamma \bar{e} }^{(0)} \otimes P_{\bar{e} e}^{(0)} 
\nonumber \\ && \quad
+ \left. \frac{1}{3} P_{ee}^{(0)} \otimes P_{ \gamma e}^{(0)} + \frac{1}{6} P_{ee}^{(0)} \otimes P_{ \gamma e}^{(0)} \otimes P_{ \gamma  \gamma }^{(0)} + \frac{1}{6} P_{ee}^{(0)} \otimes P_{ee}^{(0)} \otimes P_{ \gamma e}^{(0)} \right),
\end{eqnarray}

\begin{eqnarray}
&&  D_{e \bar{e}}^{(\mathrm{III})} (x,\mu_F^2/\mu_R^2)= D_{e \bar{e}}^{(\mathrm{II})} + \left( \frac{\alpha}{2 \pi}  \right)^3 L^2 \left( \frac{2}{3} P_{e \bar{e}}^{(1)} \right. + \frac{1}{2}  P_{\gamma \bar{e}}^{(0)} \otimes P_{e \gamma}^{(1)} 
\nonumber \\ \quad
&&  \left. - \frac{10}{9}P_{\gamma \bar{e}}^{(0)} \otimes P_{e \gamma}^{(0)}  + \frac{1}{2}  P_{\gamma \bar{e}}^{(1)} \otimes P_{e \gamma}^{(0)}  \right. + \frac{1}{3} d_{\gamma e}^{(1)} \otimes P_{e \gamma}^{(0)} - \frac{13}{54} P_{e \bar{e}}^{(0)}
\nonumber \\ \quad
&& 
 + \frac{1}{2} P_{\bar{e} \bar{e}}^{(1)} \otimes P_{e \bar{e}}^{(0)} + \frac{1}{2} P_{\bar{e} \bar{e}}^{(0)} \otimes P_{e \bar{e}}^{(1)} - \frac{10}{9} P_{\bar{e} \bar{e}}^{(0)} \otimes P_{e \bar{e}}^{(0)} + \frac{1}{2} d_{\gamma e}^{(1)} \otimes P_{\gamma \gamma}^{(0)} \otimes P_{e \gamma}^{(0)} \nonumber \\ \quad
&&  +\frac{1}{2} d_{ee}^{(1)} \otimes P_{\gamma \bar{e}}^{(0)} \otimes P_{e \gamma}^{(0)} + \frac{1}{2} P_{ee}^{(1)} \otimes P_{e \bar{e}}^{(0)} + \frac{1}{2} d_{\gamma e}^{(1)} \otimes P_{e \bar{e}}^{(0)} \otimes P_{\bar{e} \gamma}^{(0)} + \frac{1}{3} d_{ee}^{(1)} \otimes P_{e \bar{e}}^{(0)}
\nonumber \\ \quad
&& 
  + \frac{1}{2} d_{ee}^{(1)} \otimes P_{\bar{e} \bar{e}}^{(0)} \otimes P_{e \bar{e}}^{(0)} 
- \frac{10}{9}  P_{ee}^{(0)} \otimes P_{e \bar{e}}^{(0)}  + \frac{1}{2}  P_{ee}^{(0)} \otimes d_{ee}^{(1)} \otimes P_{e \bar{e}}^{(0)} 
\nonumber \\ \quad
&& 
+ \frac{1}{2} P_{ee}^{(0)} \otimes P_{e \bar{e}}^{(1)} \left. +  \frac{1}{2} P_{ee}^{(0)} \otimes d_{\gamma e}^{(1)} \otimes P_{e \gamma}^{(0)}  \right)
\nonumber \\ \quad &&
+\left( \frac{\alpha}{2 \pi}  \right)^3 L^3 \left(   \frac{4}{27} P_{e \bar{e}}^{(0)}  + \frac{1}{6} P_{\bar{e} e}^{(0)} \otimes P_{e \bar{e}}^{(0)} \otimes  P_{e \bar{e}}^{(0)} + \frac{1}{3} P_{\bar{e} \bar{e}}^{(0)} \otimes P_{e \bar{e}}^{(0)}  + \frac{1}{6} P_{\bar{e} \bar{e}}^{(0)} \otimes P_{\bar{e} \bar{e}}^{(0)} \otimes  P_{e \bar{e}}^{(0)} \right. \nonumber \\ \quad
&&  + \frac{1}{6} P_{\gamma \bar{e}}^{(0)} \otimes P_{e \bar{e}}^{(0)} \otimes P_{\bar{e} \gamma}^{(0)}  + \frac{1}{3} P_{\gamma \bar{e}}^{(0)} \otimes P_{e \gamma}^{(0)}  + \frac{1}{6} P_{\gamma \bar{e}}^{(0)} \otimes P_{\gamma \gamma}^{(0)} \otimes P_{e \gamma}^{(0)} 
\nonumber \\ \quad
&& + \frac{1}{6} P_{\gamma \bar{e}}^{(0)} \otimes P_{\bar{e} \bar{e}}^{(0)} \otimes P_{e \gamma}^{(0)}  + \frac{1}{6} P_{\gamma e}^{(0)} \otimes P_{e \gamma}^{(0)} \otimes P_{e \bar{e}}^{(0)}  + \frac{1}{3} P_{ee}^{(0)} \otimes P_{e \bar{e}}^{(0)} 
\nonumber \\ \quad
&& + \left. \frac{1}{6} P_{ee}^{(0)} \otimes P_{\bar{e} \bar{e}}^{(0)} \otimes 
         P_{e \bar{e}}^{(0)}  + \frac{1}{6} P_{ee}^{(0)} \otimes P_{\gamma \bar{e}}^{(0)} \otimes P_{e \gamma}^{(0)}  + \frac{1}{6} P_{ee}^{(0)} \otimes P_{ee}^{(0)} \otimes P_{e \bar{e}}^{(0)}    
         \right). 
\end{eqnarray}
Because of the charge conjugation parity, these functions obey the equalities (\ref{7},\ref{8},\ref{9},\ref{10}).

\section{Explicit results for PDFs}
\label{secC1}

\begin{eqnarray}
&& \bigl[ D_{ee}^{\mathrm{(I)}} (z,\mu_F^2/\mu_R^2) \bigr]_S = \delta(1-z)+ \frac{\alpha}{2 \pi} ( - 1 - 2 \ln (1-z) )  \frac{1+z^2}{1-z} +\frac{\alpha}{2 \pi} L \frac{1+z^2}{1-z}\, ,
\\  && 
\bigl[ D_{e \bar{e}}^{\mathrm{(I)}} (z,\mu_F^2/\mu_R^2) \bigr]_S = 0,
\\ &&
\bigl[ D_{\gamma e}^{\mathrm{(I)}} (z,\mu_F^2/\mu_R^2) \bigr]_S = \frac{\alpha}{2 \pi} \left( - 2 - 4 \ln z \right) \frac{(1-z)^2 + 1}{z} + \frac{\alpha}{2 \pi} L \frac{(1-z)^2 + 1}{z};
\\
&&\bigl[ D_{ee}^{(\mathrm{II})} (z,\mu_F^2/\mu_R^2) \bigr]_S =  \bigl[ D_{ee}^{\mathrm{(I)}} (z,\mu_F^2/\mu_R^2) \bigr]_S 
\nonumber \\ &&
+ \left( \frac{\alpha}{2 \pi} \right)^2 L  \Biggl[  - \frac{157}{18} - \frac{2}{z} - \frac{11}{9(1-z)} + \frac{251}{18} z - 2 z^2  + 4 \zeta_2 \frac{1+z^2}{1-z}
\nonumber \\ &&
 - ( 2  + 2 z) \mathrm{Li}_2 (1-z) + \ln z \left( - \frac{8}{3 z} - \frac{1}{3 (1-z)}  - \frac{22}{3}  - \frac{4}{3} z + \frac{8}{3} z^2 \right)
\nonumber \\ &&
+ \ln (1-z) \left( 11 - \frac{14}{1-z}  + 3 z \right) - 6 \ln^2 (1-z) \frac{1+z^2}{1-z} + \ln z \ln (1-z) \left(- 4+ \frac{4}{1-z}  - 4 z \right) 
 \nonumber \\ &&
+ \ln^2 z \left( - \frac{7}{2}   - \frac{7}{2} z \right)   \Biggr] + \left( \frac{\alpha}{2 \pi} \right)^2 L^2 \Biggl[ - \frac{7}{3} + \frac{2}{3 z} + \frac{11}{3(1-z)} - \frac{4}{3} z - \frac{2}{3} z^2  + 2 \ln (1-z) \frac{1+z^2}{1-z} 
\nonumber \\ &&
  +\ln z \left( \frac{5}{2} - \frac{2}{1-z}  + \frac{5}{2} z  \right)  \Biggr]; 
\\
&&\bigl[ D_{\gamma e}^{(\mathrm{II})} (z,\mu_F^2/\mu_R^2) \bigl]_S = \bigl[ D_{\gamma e}^{(\mathrm{I})} (z,\mu_F^2/\mu_R^2) \bigl]_S \ + \left( \frac{\alpha}{2 \pi} \right)^2 L \Biggl[  \frac{47}{18} - \frac{46}{9 z}  - \frac{2}{1-z} - \frac{133}{18} z +z^2 \nonumber \\
&&+ \ln (1-z) \left( \frac{38}{3} - \frac{38}{3 z} - \frac{4}{1-z} - \frac{25}{3} z +  2 z^2\right) 
\nonumber \\ 
&&
 - \ln^2 (1- z) \frac{(1-z)^2+1}{z} + \ln z \left( - \frac{2}{3} +\frac{8}{3 z} + \frac{29}{6} z \right) + \ln^2 z \left( -1 + \frac{z}{2}\right) \Biggr]
\nonumber \\
&& + \left( \frac{\alpha}{2 \pi} \right)^2 L^2 \Biggl[ 1  - \frac{1}{4} z + \ln (1-z) \frac{(1-z)^2 + 1}{z}  +  \ln z \left( 1 - \frac{z}{2}\right) \Biggr];
\\
&& \bigl[ D_{e \bar{e}}^{(\mathrm{II})} (z,\mu_F^2/\mu_R^2) \bigl]_S =  \left( \frac{\alpha}{2 \pi} \right)^2 L \Biggl[ 1 - \frac{2}{z} + 3 z  - 2 z^2 - 6 \zeta_2 \frac{1+z^2}{1+z} + 4 \mathrm{Li}_2 (1+z) \frac{1+z^2}{1+z}
\nonumber \\ &&
 + \ln z  \left(  - 3 - \frac{8}{3 z} + 3 z +\frac{8}{3} z^2 \right)  + \ln^2 z \left(  - 4 + \frac{2}{1+z} - 2 z \right)\Biggr]
\nonumber \\ &&
+ \left( \frac{\alpha}{2 \pi} \right)^2 L^2 \Biggl[ \frac{1}{2} + \frac{2}{3 z}  - \frac{1}{2} z - \frac{2}{3}z^2  + (1 +  z) \ln z  \Biggr], 
\end{eqnarray}

The third iteration yields
\begin{eqnarray}
&& \bigl[ D_{ee}^{(\mathrm{III})} (z,\mu_F^2/\mu_R^2) \bigr]_S = \bigl[ D_{ee}^{\mathrm{(II)}} (z,\mu_F^2/\mu_R^2) \bigr]_S
\nonumber \\ &&
+\left( \frac{\alpha}{2 \pi} \right)^3 L^2 \Biggl[  \frac{539}{27} + \frac{11}{3 z}  - \frac{37}{54 (1-z)}  - \frac{1007}{54} z - \frac{17}{3} z^2  \nonumber \\ &&
+ \zeta_2 \left( \frac{18 }{1-z} - 19      + z  + \frac{8}{3} z^2 - \frac{8}{3 z} \right) - 10 \ \zeta_3 \frac{1+z^2}{1-z}  - ( 7 + 7z)   \mathrm{S}_{1,2} (1-z) 
\nonumber \\ &&
  + \mathrm{Li}_2 (1-z)   \Bigl(  - 12 - 10 z + \frac{4}{3} z^2  +  \ln (1-z)  (  - 12 - 12 z )  
\nonumber \\ &&
 +  \ln z  (  - 9 - 9 z ) \Bigr)  + 12  (1+z) \mathrm{Li}_3 (1-z)
+  \ln (1-z)   \left(  - \frac{209}{36} - \frac{2}{3 z} - \frac{397}{18 (1-z)}  \right.
\nonumber \\ &&
 \left. + \frac{1075}{36} z  - \frac{4}{3} z^2  +   20 \zeta_2 \frac{1+z^2}{1-z}\right) +  \ln^2 (1-z)   \left( \frac{53}{2} - \frac{2}{3 z} - \frac{34}{1-z} + \frac{15}{2} z + \frac{2}{3} z^2 \right) 
\nonumber \\ &&
 - 8 \ln^3 (1-z) \ \frac{1+z^2}{1-z} +  \ln z   \left( \frac{65}{4} + \frac{17}{6 (1-z)}  - \frac{61}{4}z + \frac{28}{9} z^2 - 4 \zeta_2 \frac{1+z^2}{1-z}\right) 
\nonumber \\ &&
 +  \ln z \ln (1-z)   \left(  - \frac{68}{3} - \frac{8}{3 z} + \frac{28}{3 (1-z)} - \frac{26}{3} z + 4 z^2 \right) +  12 \ln z \ln^2 (1-z) \frac{z^2}{1-z}  
\nonumber \\ &&
   +  \ln^2 z   \left(  - \frac{1}{2} + \frac{7}{3 (1-z)} - \frac{7}{2} z - 4 z^2 \right)  +  \ln^2 z \ln (1-z)   \left(  - \frac{11}{2} - \frac{11}{2} z \right)   +  \ln^3 z   \left( \frac{4}{3} + \frac{4}{3} z \right)  \Biggr] 
\nonumber \\ &&  
  + \left( \frac{\alpha}{2 \pi} \right)^3 L^3 \Biggl[  - \frac{971}{216} + \frac{8}{27 z} + \frac{491}{108 (1-z)} - 2 \zeta_2 \frac{1+z^2 }{1-z} - \frac{11}{216} z  - \frac{8}{27} z^2 
\nonumber \\ &&  
+  \mathrm{Li}_2 (1-z)   \left( \frac{7}{3} + \frac{7}{3}z \right)  +  \ln (1-z)   \left(  - \frac{17}{3} + \frac{8}{9 z} + \frac{26}{3 (1-z)} - 3 z - \frac{8}{9} z^2 \right)
\nonumber \\ &&  
 +  2 \ln^2 (1-z) \frac{1+z^2 }{1-z}  +  \ln z   \left( \frac{151}{36} - \frac{13}{3 (1-z)} + \frac{103}{36} z + \frac{8}{9} z^2 \right) 
\nonumber \\ &&
+ \ln z \ln (1-z)   \left(  \frac{13}{3} - \frac{4}{1-z}  + \frac{13}{3} z \right)+  \ln^2 z   \left( - \frac{11}{12} + \frac{2}{3 (1-z)} - \frac{11}{12} z\right) \Biggr],
\end{eqnarray}

\begin{eqnarray}
&& \bigl[ D_{\gamma e}^{(\mathrm{III})} (z,\mu_F^2/\mu_R^2) \bigr]_S = \bigl[ D_{\gamma e}^{(\mathrm{II})} (z,\mu_F^2/\mu_R^2) \bigr]_S
+\left( \frac{\alpha}{2 \pi} \right)^3 L^2   \Biggl[ -\frac{4879}{108}  + \frac{55}{18 z} 
- \frac{2}{3 (1-z)} 
\nonumber \\ &&
+ \frac{14129}{432} z  
 + \frac{37}{27} z^2 +   \left( - \frac{40}{3}  - \frac{23}{6}z - \frac{20}{3 z} \right)\zeta_2     
+ \left( 14 +  z\right)\zeta_3  
+ \left(- 10- \frac{4}{z} + z \right) \mathrm{S}_{1,2} (1-z) 
\nonumber \\ && + \mathrm{Li}_2 (1+z) \left( 10 + \frac{6}{z} + 4 z + \frac{4}{z} \right) 
  + \mathrm{Li}_2 (1-z) \biggl( - \frac{16}{3} - \frac{16}{3 z}  - \frac{5}{3}z 
 \nonumber \\ && 
  + \ln (1-z) \left( -6 -\frac{8}{z} + 3z\right)   + \ln z \left( - 8 + \frac{4}{z} \right) \biggr) 
 \nonumber \\ &&  
+ \mathrm{Li}_3(1 - z)   \left( 6 + \frac{8}{z} - 4 z \right)   + \mathrm{Li}_3(z^2) \left(  - 1 + \frac{1}{z}  - \frac{z}{2}\right) +   \mathrm{Li}_3 (1+z) \left(  -8 - \frac{8}{z} - 4 z\right)
 \nonumber \\ &&  
 + \ln (1+z) \zeta_2 \left( 12 + \frac{12}{z} + 6 z \right) 
 + \ln (1-z)   \left( \frac{95}{9} - \frac{253}{18 z} - \frac{4}{3 (1-z)} - \frac{487}{36} z + \frac{2}{3} z^2 \right.
 \nonumber \\ &&  
 \left.- 6 \frac{(1-z)^2 + 1}{z} \zeta_2 \right) 
 +  \ln^2 (1-z)   \left(  - 1 + \frac{6}{ z} - \frac{13}{4}z \right)
\nonumber \\ && 
 - 3  \ln^3 (1-z) \frac{(1-z)^2 + 1}{z} +  \ln z   \left(  - \frac{137}{4}  + \frac{20}{3 z} + \frac{11}{2 (1-z)} - \frac{23}{12} z   \right.
\nonumber \\ && 
\left. - \frac{127}{36} z^2 + \left(6 - \frac{8}{z} - z \right) \zeta_2 + \frac{4}{z} \mathrm{Li}_2 (1+z) \right)  + \ln z \ln (1-z)   \left( \frac{7}{3} - \frac{19}{6} z  \right)  
\nonumber \\ && 
+  \ln z \ln^2 (1-z)  \left( -7 + \frac{7}{2} z \right) 
+  \ln^2 z   \left(  - \frac{11}{3} + \frac{8}{3 z} - \frac{2}{3}\right) 
\nonumber \\ &&  
  +   \ln (1+z) (- 2 +\frac{2}{z}) +  \ln (1-z)  \left( -3 +\frac{2}{z} - \frac{1}{2} \right) 
  + \ln^3 z   \left(  - \frac{7}{3} + \frac{3}{2} z \right)  \Biggr]   
\nonumber \\ &&  
+ \left( \frac{\alpha}{2 \pi} \right)^3 L^3   \Biggl[ \left( \frac{193}{36} - \frac{62}{27 z} - \frac{191}{72} z + \frac{8}{27} z^2 \right)  - \frac{2}{3} \zeta_2   \frac{(1-z)^2 + 1}{z}  +  \frac{4}{3 z}  \mathrm{Li}_2 (1-z)   
\nonumber \\ &&  
+  \ln (1-z)   \left( \frac{4}{9} + \frac{8}{9 z} + \frac{1}{9} z \right) 
+  \frac{2}{3} \ln^2 (1-z)   \frac{(1-z)^2 + 1}{z}  +  \ln z   \left(  - \frac{7}{18} - \frac{8}{9 z} + \frac{10}{9} z \right)
\nonumber \\ &&  
 +  \ln z \ln (1-z)   \left( \frac{4}{3} - \frac{2}{3} z \right)
 +  \ln^2 z   \left( \frac{1}{2} - \frac{1}{4} z \right)   \Biggr],
\end{eqnarray}


\begin{eqnarray}
&& \bigl[ D_{e \bar{e}}^{(\mathrm{III})} (z,\mu_F^2/\mu_R^2) \bigr]_S = \bigl[ D_{e \bar{e}}^{(\mathrm{II})} (z,\mu_F^2/\mu_R^2) \bigr]_S 
 +  \left( \frac{\alpha}{2 \pi} \right)^3 L^2   \Biggl[  \frac{137}{36} + \frac{11}{3 z} - \frac{65}{36} z - \frac{17}{3} z^2 
 \nonumber \\ &&
+  \mathrm{Li}_2 (1+z) \left( \frac{52}{3 (1+z)} - \frac{50}{3} + \frac{2}{3} z \right)
 +\zeta_3 \left( 8 - \frac{12}{1+z}  - 8 z \right)
\nonumber \\ &&
+ \zeta_2 \left( 23 - \frac{8}{3 z} - \frac{26}{1+z}  + z + \frac{8}{3} z^2 \right) + \left( - 10  + 6 z \right) \mathrm{S}_{1,2} (1-z)
\nonumber \\ &&
 + \mathrm{Li}_2 (1-z) \Biggl\{ -\frac{4}{3} + \frac{2}{3} z + \frac{4}{3} z^2 -  8 \ln (1+z) \frac{1+z^2}{1+z}  +  \ln (1-z) \Bigl(  - 10 + \frac{16}{1+z} 
  \nonumber \\ &&  
 + 6 z \Bigr)+ (  - 14 + 2 z )  \ln z \Biggr\} - 8  \left( \mathrm{Li}_2\left(\frac{1+z}{2}\right)  -\mathrm{Li}_2 \left(\frac{1-z}{1+z}\right) \right) \bigl( \ln (1+z)  
 \nonumber \\ && 
 - \ln(1- z) \bigr) \frac{1+z^2}{1+z} + \mathrm{Li}_3 (1-z) (- 6 + \frac{16}{1+z} + 10 z) 
 + 16 \mathrm{Li}_3(1 + z) \frac{1+z^2}{1+z} 
  \nonumber \\ &&  
- 4 \ \mathrm{Li}_3(1- z^2) \frac{1+z^2}{1+z} + \mathrm{Li}_3(z^2) (-1 - \frac{2}{1+z}+  z)
- 8  \mathrm{Li}_3 \left(\frac{ z}{1 + z} \right) \frac{1+z^2}{1+z} 
\nonumber \\
&& 
 + \ln (1+z) \left( - 32 \zeta_2 -  4  \ln^2 (2)  + 8 \mathrm{Li}_2 (1+z)\right) \frac{1+z^2}{1+z}
+ \frac{16}{3} \ln (1+z)^3 \frac{1+z^2}{1+z} 
\nonumber \\ && 
+ \ln (1-z)   \left( \frac{1}{2} - \frac{2}{3 z} + \frac{3}{2} z   - \frac{4}{3} z^2  + (- 8 \zeta_2 + 4 \ln^2 (2) ) \frac{1+z^2}{1+z}\right) 
\nonumber \\ && 
 + 8 \ln (1-z) \ln (1+z) \ln (2) \frac{1+z^2}{1+z}  - 12 \ln (1-z) \ln^2 (1+z) \frac{1+z^2}{1+z}
\nonumber \\ && 
 +  \ln^2 (1-z)   \left(  - \frac{1}{2} - \frac{2}{3 z} + \frac{1}{2} z + \frac{2}{3} z^2  - 8 \ln (2) \frac{1+z^2}{1+z} \right)  
\nonumber \\ && 
 + 8 \ln^2 (1-z) \ln (1+z) \frac{1+z^2}{1+z}   + \ln z   \Bigl( \frac{433}{36}  + \frac{151}{36} z  + \frac{28}{9} z^2  + ( -6 + \frac{24}{1+z} - 2 z) \zeta_2
\nonumber \\ &&
 - 8 \frac{\mathrm{Li}_2 (1+z)}{1+z} \Bigr) - 4 \ln z \ln^2 (1+z) \frac{1+z^2}{1+z}  +  \ln z \ln (1-z)  
\nonumber \\ &&
   \times \left(  - \frac{10}{3} - \frac{8}{3 z} + \frac{8}{3} z + 4 z^2 \right) 
+   (  - 1 - z ) \ln z \ln^2 (1-z)   +  \ln^2 z   \left(  - \frac{10}{3} \right. + \frac{13}{3 (1+z)}
\nonumber \\ && 
\left.  - \frac{9}{2} z - 4 z^2 \right) + \ln^2 z \ln (1+z) \left( - 6 + \frac{4}{1+z} + 6 z \right)   + \ln^2 z \ln (1-z) \left(    - 9 - z \right)
 \nonumber \\ &&  
 + \ln^3 z  \left(\frac{13}{6} - \frac{4}{3 (1+z)} + \frac{1}{6} z \right)
  \Biggr] +  \left( \frac{\alpha}{2 \pi} \right)^3 L^3   \Biggl[ - \frac{19}{18} + \frac{8}{27} 	 + \frac{19}{18} z - \frac{8}{27} z^2
	\nonumber \\ && 
	+ \left( \frac{4}{3} z  + \frac{4}{3} \right)  \mathrm{Li}_2 (1-z) + \ln (1- z)  \left( \frac{8}{9z} + \frac{2}{3}  - \frac{2}{3} z - \frac{8}{9} z^2  \right) 
	\nonumber \\ && 
	+ \ln z \left( \frac{4}{9} + \frac{10}{9} z + \frac{8}{9} z^2 \right) + \ln z \ln (1- z) \left( \frac{4}{3}  + \frac{4}{3} z \right)
	 + \left( - \frac{1}{3}  - \frac{1}{3} z \right) \ln^2 z   \Biggr].
\end{eqnarray}

And for timelike functions we get
\begin{eqnarray}
&& \bigl[ D_{ee}^{\mathrm{(I)}} (z,\mu_F^2/\mu_R^2) \bigr]_T = \bigl[ D_{ee}^{\mathrm{(I)}} (z,\mu_F^2/\mu_R^2) \bigr]_S,  \\
&&  \bigl[ D_{e \bar{e}}^{\mathrm{(I)}} (z,\mu_F^2/\mu_R^2) \bigr]_T = 0 = \bigl[ D_{e \bar{e}}^{\mathrm{(I)}} \bigr]_S, \\
&& \bigl[ D_{\gamma e}^{\mathrm{(I)}} (z,\mu_F^2/\mu_R^2) \bigr]_T = \bigl[ D_{\gamma e}^{\mathrm{(I)}} (z,\mu_F^2/\mu_R^2) \bigr]_S, \\
&& \bigl[ D_{ee}^{(\mathrm{II})} (z,\mu_F^2/\mu_R^2) \bigr]_T = \bigl[ D_{ee}^{(\mathrm{I})} (z,\mu_F^2/\mu_R^2) \bigr]_T + \left( \frac{\alpha}{2 \pi} \right)^2 L \Biggl[  
- \frac{265}{18} - \frac{58}{3 z} - \frac{11}{9 (1-z)} + \frac{215}{18} z 
\nonumber \\ &&
+ \frac{94}{9} z^2 + 4 \frac{1+z^2}{1-z} \zeta_2  
+  \mathrm{Li}_2 (1-z)   (  - 2 - 2 z ) +  \ln (1-z)   \left( 11 - \frac{14}{1-z} + 3 z \right) 
\nonumber \\ &&
- 6 \ln^2 (1-z)  \frac{1+z^2}{1-z} 
+  \ln z   \left(  - \frac{55}{3} - \frac{8}{3 z} + \frac{17}{3 (1-z)} - \frac{49}{3} z - \frac{8}{3} z^2 \right) 
\nonumber \\ &&
+  \ln z \ln (1-z)   \left(  - 8 + \frac{12}{1-z} - 8 z \right)
+  \ln^2 z   \left( \frac{3}{2} - \frac{4}{1-z} + \frac{3}{2} z \right) \Biggr] + \left( \frac{\alpha}{2 \pi} \right)^2 L^2 \Biggl[ - \frac{7}{3} + \frac{2}{3 z} 
\nonumber \\ &&
+ \frac{11}{3 (1-z)} - \frac{4}{3} z - \frac{2}{3} z^2 
+  2 \ln (1-z)  \frac{1+z^2}{1-z} +  \ln z   \left( \frac{5}{2} - \frac{2}{1-z} + \frac{5}{2} z \right)  \Biggr],
\\ &&
\bigl[ D_{\gamma e}^{\mathrm{(II)}} (z,\mu_F^2/\mu_R^2) \bigr]_T = \bigl[ D_{\gamma e}^{\mathrm{(I)}} (z,\mu_F^2/\mu_R^2) \bigr]_T
+ \left( \frac{\alpha}{2 \pi} \right)^2 L \Biggl[  \frac{1}{6} - \frac{2}{3 z} - \frac{2}{1-z} + \frac{25}{6} z + z^2 
\nonumber \\ &&
- 8 \zeta_2 \frac{(1-z)^2 +1}{z} 
- 8 \mathrm{Li}_2 (1-z) \frac{(1-z)^2 +1}{z}  +  \ln (1-z)   \left( 4 - \frac{4}{z} - \frac{4}{1-z} + 2 z^2 \right) 
\nonumber \\ &&
+  \ln^2 (1-z)  \frac{(1-z)^2 +1}{z}  
+  \ln z   \left(  - \frac{32}{3} + \frac{8}{3 z} + \frac{11}{6} z \right)   +  4 \ln z \ln (1-z) \frac{(1-z)^2 +1}{z}  
\nonumber \\ && 
+  \ln^2 z   \left( 1 - \frac{1}{2} z \right)   \Biggr] 
+ \left( \frac{\alpha}{2 \pi} \right)^2 L^2 \Biggl[ 1  - \frac{1}{4} z + \ln z\left( 1 - \frac{1}{2}z \right) + \ln (1-z) \frac{(1-z)^2 +1}{z} \Biggr],
\\ &&
\bigl[ D_{e \bar{e}}^{\mathrm{(II)}} (z,\mu_F^2/\mu_R^2) \bigr]_T =  \left( \frac{\alpha}{2 \pi} \right)^2 L  \Biggl[    1 - \frac{2}{z} + 3 z 
- 2 z^2 - 6 \zeta_2 \frac{1+z^2}{1+z}  + 4 \frac{1+z^2}{1+z} \mathrm{Li}_2 (1+z)
\nonumber \\ &&
+  \ln z   \left(  - 3 - \frac{8}{3 z} + 3 z + \frac{8}{3} z^2 \right)    +  \ln^2 z   \left(  - 4 + \frac{2}{1+z} - 2 z \right) \Biggr] \nonumber \\ &&
+ \left( \frac{\alpha}{2 \pi} \right)^2 L^2   \Biggl[ \frac{1}{2} + \frac{2}{3 z} - \frac{1}{2} z -\frac{2}{3} z^2 + (1 + z) \ln z  \Biggr] ,
\end{eqnarray}

\begin{eqnarray}
&& \bigl[ D_{e e}^{(\mathrm{III})} (z,\mu_F^2/\mu_R^2)\bigr]_T = \bigl[ D_{ee}^{(\mathrm{II})} (z,\mu_F^2/\mu_R^2) \bigr]_T + \left( \frac{\alpha}{2 \pi} \right)^3 L^2   \Biggl[    \frac{917}{27} - \frac{62}{27 z} + \frac{37}{54 (1-z)} 
\nonumber \\ && 
  - \frac{1709}{54} z - \frac{19}{29} z^2  - 10 \zeta_3 \frac{1+z^2 }{1-z} + \zeta_2 \left( \frac{8}{3 z}   + \frac{18}{1-z} - 15 - 3z  - \frac{8}{3} z^2 \right)
\nonumber \\ &&
+   \mathrm{S}_{1,2} (1-z)   (  - 51 - 51 z ) + \mathrm{Li}_2 (1-z)   \left(  - \frac{103}{3} - \frac{16}{z} - \frac{40}{3} z + 12 z^2  \right.
\nonumber \\ &&
\left.  - (14 \ln (1-z) + 21  \ln z )  (  1+z )  \right) +   \mathrm{Li}_3 (1-z)   ( 14 + 14 z ) \nonumber \\
&&+   \ln (1-z)   \left(  - \frac{527}{36} - \frac{49}{9 z} - \frac{397}{18 (1-z)} + \frac{1213}{36} z + \frac{76}{9} z^2 \right.
\nonumber \\ &&
\left. + 20 \zeta_2 \frac{1+z^2}{1-z}\right) +   \ln^2 (1-z)   \left( 27 -  \frac{2}{z} - \frac{34 }{1-z} + 7 z + 2 z^2 \right)  - 8   \ln^3 (1-z) \frac{1+z^2}{1-z} 
\nonumber \\ &&
+   \ln z   \left( \frac{31}{36} + \frac{205}{18 (1-z)} - \frac{545}{36} z - \frac{20}{3} z^2 + 20 \zeta_2 \frac{1+z^2}{1-z} \right)  
\nonumber \\ &&
+   \ln z \ln (1-z)   \left(  - \frac{161}{3} + \frac{116}{3 (1-z)} - \frac{98}{3} z - 4 z^2 \right) 
\nonumber \\ &&
+   \ln z \ln^2 (1-z)   \left(  - 21 + \frac{28}{1-z} - 21 z \right) +   \ln^2 z   \left( \frac{43}{2} - \frac{37}{3 (1-z)} + \frac{37}{2} z + 8 z^2 \right) \nonumber \\ &&
+   \ln^2 z \ln (1-z)   \left( \frac{25}{2} - \frac{16}{1-z} + \frac{25}{2} z \right)  +   \ln^3 z   \left(  - 3 + \frac{8}{3 (1-z)} - 3 z \right) \Biggr] 
\nonumber \\ && 
+ \left( \frac{\alpha}{2 \pi} \right)^3 L^3   \Biggl[         - \frac{971}{216} + \frac{8}{27 z} + \frac{491}{108 (1-z) } - \frac{11}{216} z - \frac{8}{27}z^2  - 2 \zeta_2 \frac{1+z^2}{1-z} 
\nonumber \\ && 
+  \mathrm{Li}_2 (1-z)   \left( \frac{7}{3} + \frac{7}{3} z \right) +  \ln (1-z)   \left(  - \frac{17}{3} + \frac{8}{9 z} + \frac{26}{3 (1-z)} - 3 z - \frac{8}{9} z^2 \right)  
\nonumber \\ && 
+ 2  \ln^2 (1-z) \frac{1+z^2}{1-z}+  \ln z   \left( \frac{151}{36} - \frac{13}{3 (1-z)} + \frac{103}{36} z + \frac{8}{9} z^2 \right) 
\nonumber \\ && 
+  \ln z \ln (1-z)   \left( \frac{13}{3} - \frac{4}{1-z} + \frac{13}{3} z \right) +  \ln^2 z   \left(  - \frac{11}{12} + \frac{2}{3 (1-z)} - \frac{11}{12} z \right) \Biggr] ,
\end{eqnarray}


\begin{eqnarray}
&& \bigl[ D_{\gamma e}^{(\mathrm{III})} (z,\mu_F^2/\mu_R^2) \bigr]_T  = \bigl[ D_{\gamma e}^{(\mathrm{II})} (z,\mu_F^2/\mu_R^2) \bigr]_T 
\nonumber \\ &&  
 + \left( \frac{\alpha}{2 \pi} \right)^3 L^2   \Biggl[ - \frac{557}{24} - \frac{19}{54 z} + \frac{10}{3 (1-z)} + \frac{233}{16} z  + \frac{353}{108} z^2 
\nonumber \\ &&        
+ \zeta_3    \left( 10 + \frac{4}{z} + 3 z \right)+ \zeta_2    \left( - \frac{70}{3 z} - \frac{8}{1-z} + \frac{10}{3} - \frac{85}{6} z + 4 z^2 \right)  +  \mathrm{S}_{1,2} (1-z)   \left(  - 4 - \frac{4}{z} - 2 z\right)
 \nonumber \\ && 
+  \mathrm{Li}_2 (1+z) \left( \frac{6}{z} + 10  + 4z  + \frac{4}{z}\ln z \right) +  \mathrm{Li}_2 (1-z)   \left( 7 - \frac{14}{3 z} - \frac{8}{1-z} - 6 z + 4 z^2   \right.
\nonumber \\ &&       
 \left.  + \ln z \left( - 6 + \frac{4}{z} - z  \right)- \frac{12}{z}  \ln (1-z) \right) 
 +  \frac{12}{z} \mathrm{Li}_3 (1-z)
 +  \mathrm{Li}_3(1 + z) \left( - 8 - \frac{8}{z} - 4 z\right) 
\nonumber \\ &&  
+  \mathrm{Li}_3(z^2)   (  - 1 + \frac{1}{z}- \frac{1}{2}z )  +   \ln (1+z)   \zeta_2 \left(  12 + \frac{12}{z} + 6 z \right) 
  +  \ln (1-z)   \left( \frac{5}{6} - \frac{1}{2 z}\right.
 \nonumber \\ && 
  \left.  + \frac{2}{3 (1-z)} - \frac{23}{6} z - \frac{1}{3} z^2  + 4 \zeta_2  \frac{(1-z)^2 + 1}{z}\right) +  \ln^2 (1-z)   \left(  - \frac{31}{3} + \frac{13}{3 z} + \frac{1}{1-z} \right. 
  \nonumber \\ &&
  \left. + \frac{11}{3} z - \frac{1}{2} z^2 \right) -2 \ln^3 (1-z)   \frac{(1-z)^2 + 1}{z}+  \ln z   \left(  - \frac{35}{3} - \frac{16}{9 z} - \frac{15}{2 (1-z)} - \frac{13}{4} z \right. 
 \nonumber \\ && 
 \left. - \frac{73}{36}z^2  + \zeta_2 \left( 6 - \frac{8}{z} -  z \right) \right) +  \ln z \ln (1-z)   \left( \frac{13}{3} + \frac{4}{z} + \frac{4 }{1-z} - \frac{19}{6} z - 2 z^2 \right)  
   \nonumber \\ &&
 +  \ln z \ln^2 (1-z)   (  - 6 + 3 z ) 
 +  \ln^2 z   \biggl( \frac{43}{12}  + \frac{1}{2 (1-z)} + \frac{17}{12} z + \frac{1}{4} z^2  
   \nonumber \\ &&
 + \ln (1+z) \left( -2 + \frac{2}{z} - z \right)+ \ln (1-z)   \left( \frac{2}{z} - 2 z \right) \biggr) 
\nonumber \\ &&
+  \ln^3 z   \left( - \frac{5}{6} + \frac{3}{4} z \right)   \Biggr] + \left( \frac{\alpha}{2 \pi} \right)^3 L^3  \Biggl[  \frac{193}{36} - \frac{62}{27 z} - \frac{191}{72} z  + \frac{8}{27} z^2 -  \frac{2}{3} \frac{(1-z)^2+1}{z}\zeta_2 
\nonumber \\ &&
+ \frac{4}{3 z}   \mathrm{Li}_2 (1-z)   
+ \ln (1-z)   \left( \frac{4}{9} + \frac{8}{9 z} + \frac{1}{9} z \right) + \frac{2}{3} \ln^2 (1-z) \frac{(1-z)^2+1}{z} 
\nonumber \\ &&
+ \ln z   \left(  - \frac{7}{18} - \frac{8}{9 z} + \frac{10}{9} z \right) + \frac{2}{3}\ln z \ln (1-z)   ( 2 -  z )  + \frac{1}{4} \ln^2 z   \left( 2 -  z \right) \Biggr],
\end{eqnarray}

\begin{eqnarray}
&&  \bigl[ D_{e \bar{e}}^{(\mathrm{III})} (z,\mu_F^2/\mu_R^2) \bigr]_T =   \bigl[ D_{e \bar{e}}^{(\mathrm{II})} (z,\mu_F^2/\mu_R^2) \bigr]_T 
 + \left( \frac{\alpha}{2 \pi} \right)^3 L^2  \Biggl[ \frac{109}{36} - \frac{2}{27 z}  - \frac{97}{36} z - \frac{7}{27} z^2 
\nonumber \\ && 
+\zeta_3 \left( 8 - \frac{12}{1+z}  - 8 z \right) + \zeta_2  \left( 25 +   \frac{8}{3 z} - \frac{22}{1+z} -  z - \frac{8}{3} z^2 \right) 
\nonumber \\ && 
 + \mathrm{S}_{1,2} (1-z) \left( - 62 - 46 z\right) + \mathrm{Li}_{2} (1+z) \left( - \frac{46}{3} + \frac{44}{3(1+z)} - \frac{2}{3} z\right)
\nonumber \\ && 
+ \mathrm{Li}_{2} (1-z) \Biggl(  - \frac{35}{3} - \frac{16}{z} + \frac{76}{3} z + \frac{68}{3} z^2 - 8 \ln (1+z) \frac{1+ z^2}{1 + z} +\ln (1-z) \left(  - 12 + \frac{16}{1+z} + 4 z \right)
\nonumber \\ && 
  + \ln z (  - 38 - 22 z )  \Biggr)
- 8 \left(\mathrm{Li}_2 \Bigl(\frac{1 +z}{2} \Bigr) - \mathrm{Li}_2 \Bigl(\frac{1-z}{1 + z}\Bigr) \right)  (  \ln (1+z)  -   \ln (1-z) ) \frac{1+z^2}{1+z}
\nonumber \\ && 
+ \mathrm{Li}_{3} (1-z) (- 4 + \frac{16}{1+z}+ 12 z) + 16 \ \mathrm{Li}_3(1 + z) \frac{1+z^2}{1-z} - 4 \ \mathrm{Li}_3(1- z^2) \frac{1+z^2}{1+z}
\nonumber \\ && 
  +  \mathrm{Li}_3( z^2) \left(  - 1 - \frac{2}{1+z} + z \right) - 8 \mathrm{Li}_3 \Bigl(\frac{z}{1+z} \Bigr) \frac{1+z^2}{1+z}
+   \ln (1+z)   \left( 8 \mathrm{Li}_2 (1+z) - 32 \zeta_2    \right.
\nonumber \\ && 
\left. - 4 \ln^2 (2)   \right) \frac{1+z^2}{1+z}+  \frac{16}{3} \ln (1+z)^3  \frac{1+z^2}{1+z}  +   \ln (1-z)   \left( \frac{11}{3} + \frac{31}{9 z} + \frac{28}{3} z - \frac{148}{9} z^2 \right.
\nonumber \\ && 
\left.   + \left( -  8 \zeta_2  + 4 \ln^2 (2) \right)    \frac{1+z^2}{1+z} \right)  + 8 \ln (1-z) \ln (1+z) \ln (2) \frac{1+z^2}{1+z}  
\nonumber \\ && 
- 12 \ln (1-z) \ln^2 (1+z) \frac{1+z^2}{1+z}
+   \ln^2 (1-z)   \left(  -  \frac{2}{z} + 2 z^2  - 8 \frac{1+z^2}{1+z} \ln (2) \right) 
\nonumber \\ && 
+ 8 \ln^2 (1-z) \ln (1+z) \frac{1+z^2}{1+z}
+ \ln z   \left( \frac{59}{12} + \frac{119}{12} z + \frac{188}{9} z^2 + \zeta_2 \biggl( 2 + \frac{24}{1+z} + 6 z \right)  
\nonumber \\ && 
-  \frac{8 \mathrm{Li}_2 (1+z)}{1+z} \biggr)
- 4\ln z \ln^2 (1+z) \frac{1+z^2}{1+z}
 +   \ln z \ln (1-z)   \left( \frac{1}{3} + \frac{40}{3} z + \frac{20}{3} z^2  \right) 
 \nonumber \\ && 
 - 2(1+z) \ln z \ln^2 (1-z) +   \ln^2 z   \biggl( 1 + \frac{11}{3} z  + \frac{19}{2} z 
 + \frac{8}{3} z^2  +  \ln (1+z) \left( \frac{4}{1+z} + 6z  -6 \right)
 \nonumber \\ &&
  +   \ln (1-z)   \left(  - 7+  z \right) \biggr) - \ln^3 z   \left(  \frac{5}{6} - \frac{4}{3 (1+z)} + \frac{7}{6} z \right) \Biggr]
\nonumber \\ &&
 + \left( \frac{\alpha}{2 \pi} \right)^3 L^3   \Biggl[   - \frac{19}{18} + \frac{8}{27 z} + \frac{19}{18} z - \frac{8}{27} z^2  
 + \frac{4}{3} \mathrm{Li}_2 (1-z)   ( 1 + z)   + \ln (1-z)   \biggl( \frac{2}{3} + \frac{8}{9 z} 
 \nonumber \\ && 
  - \frac{2}{3} z - \frac{8}{9} z^2 \biggr) 
 + \ln z   \left( \frac{4}{9} + \frac{10}{9} z + \frac{8}{9} z^2 \right) + \frac{4}{3}   ( 1 +  z )\ln z \ln (1-z)  
 - \frac{1}{3} \ln^2 z   (  1 + z ) \Biggr].
\end{eqnarray}


\section*{References}

\bibliographystyle{iopart-num}
\bibliography{QED_PDF_arxiv_v2}

\providecommand{\newblock}{}
\begin{thebibliography}{10}
\expandafter\ifx\csname url\endcsname\relax
  \def\url#1{{\tt #1}}\fi
\expandafter\ifx\csname urlprefix\endcsname\relax\def\urlprefix{URL }\fi
\providecommand{\eprint}[2][]{\url{#2}}

\bibitem{FCC:2018evy}
Abada A {\em et~al.\/} (FCC) 2019 {\em Eur. Phys. J. ST\/} {\bf 228} 261--623

\bibitem{CEPCStudyGroup:2018ghi}
Dong M {\em et~al.\/} (CEPC Study Group) 2018  (\textit{Preprint}
  \eprint{1811.10545})

\bibitem{Jadach:2019bye}
Jadach S and Skrzypek M 2019 {\em Eur. Phys. J. C\/} {\bf 79} 756
  (\textit{Preprint} \eprint{1903.09895})

\bibitem{Kuraev:1985hb}
Kuraev E~A and Fadin V~S 1985 {\em Sov. J. Nucl. Phys.\/} {\bf 41} 466--472

\bibitem{DeRujula:1979grv}
De~Rujula A, Petronzio R and Savoy-Navarro A 1979 {\em Nucl. Phys. B\/} {\bf
  154} 394--426

\bibitem{Gribov:1972ri}
Gribov V~N and Lipatov L~N 1972 {\em Sov. J. Nucl. Phys.\/} {\bf 15} 438--450

\bibitem{Altarelli:1977zs}
Altarelli G and Parisi G 1977 {\em Nucl. Phys. B\/} {\bf 126} 298--318

\bibitem{Dokshitzer:1977sg}
Dokshitzer Y~L 1977 {\em Sov. Phys. JETP\/} {\bf 46} 641--653

\bibitem{Cacciari:1992pz}
Cacciari M, Deandrea A, Montagna G and Nicrosini O 1992 {\em EPL\/} {\bf 17}
  123--128

\bibitem{Yennie:1961ad}
Yennie D~R, Frautschi S~C and Suura H 1961 {\em Annals Phys.\/} {\bf 13}
  379--452

\bibitem{Przybycien:1992qe}
Przybycien M 1993 {\em Acta Phys. Polon. B\/} {\bf 24} 1105--1114
  (\textit{Preprint} \eprint{hep-th/9511029})

\bibitem{Arbuzov:1999cq}
Arbuzov A~B 1999 {\em Phys. Lett. B\/} {\bf 470} 252--258 (\textit{Preprint}
  \eprint{hep-ph/9908361})

\bibitem{Blumlein:2004bs}
Blumlein J and Kawamura H 2005 {\em Nucl. Phys. B\/} {\bf 708} 467--510
  (\textit{Preprint} \eprint{hep-ph/0409289})

\bibitem{Berends:1987ab}
Berends F~A, van Neerven W~L and Burgers G~J~H 1988 {\em Nucl. Phys. B\/} {\bf
  297} 429 [Erratum: Nucl.Phys.B 304, 921 (1988)]

\bibitem{Arbuzov:2002cn}
Arbuzov A and Melnikov K 2002 {\em Phys. Rev. D\/} {\bf 66} 093003
  (\textit{Preprint} \eprint{hep-ph/0205172})

\bibitem{Blumlein:2002fy}
Blumlein J and Kawamura H 2003 {\em Phys. Lett. B\/} {\bf 553} 242--250
  (\textit{Preprint} \eprint{hep-ph/0211191})

\bibitem{Blumlein:2007kx}
Blumlein J and Kawamura H 2007 {\em Eur. Phys. J. C\/} {\bf 51} 317--333
  (\textit{Preprint} \eprint{hep-ph/0701019})

\bibitem{Blumlein:2011mi}
Blumlein J, De~Freitas A and van Neerven W 2012 {\em Nucl. Phys. B\/} {\bf 855}
  508--569 (\textit{Preprint} \eprint{1107.4638})

\bibitem{Ablinger:2020qvo}
Ablinger J, Bl\"umlein J, De~Freitas A and Sch\"onwald K 2020 {\em Nucl. Phys.
  B\/} {\bf 955} 115045 (\textit{Preprint} \eprint{2004.04287})

\bibitem{Blumlein:2021jdl}
Bl\"umlein J, De~Freitas A and Sch\"onwald K 2021 {\em Phys. Lett. B\/} {\bf
  816} 136250 (\textit{Preprint} \eprint{2102.12237})

\bibitem{Frixione:2019lga}
Frixione S 2019 {\em JHEP\/} {\bf 11} 158 (\textit{Preprint}
  \eprint{1909.03886})

\bibitem{Bertone:2019hks}
Bertone V, Cacciari M, Frixione S and Stagnitto G 2020 {\em JHEP\/} {\bf 03}
  135 [Erratum: JHEP 08, 108 (2022)] (\textit{Preprint} \eprint{1911.12040})

\bibitem{Frixione:2012wtz}
Frixione S 2021 {\em JHEP\/} {\bf 07} 180 [Erratum: JHEP 12, 196 (2012)]
  (\textit{Preprint} \eprint{2105.06688})

\bibitem{Bertone:2022ktl}
Bertone V, Cacciari M, Frixione S, Stagnitto G, Zaro M and Zhao X 2022 {\em
  JHEP\/} {\bf 10} 089 (\textit{Preprint} \eprint{2207.03265})

\bibitem{Spiesberger:1994dm}
Spiesberger H 1995 {\em Phys. Rev. D\/} {\bf 52} 4936--4940 (\textit{Preprint}
  \eprint{hep-ph/9412286})

\bibitem{Roth:2004ti}
Roth M and Weinzierl S 2004 {\em Phys. Lett. B\/} {\bf 590} 190--198
  (\textit{Preprint} \eprint{hep-ph/0403200})

\bibitem{Martin:2004dh}
Martin A~D, Roberts R~G, Stirling W~J and Thorne R~S 2005 {\em Eur. Phys. J.
  C\/} {\bf 39} 155--161 (\textit{Preprint} \eprint{hep-ph/0411040})

\bibitem{Ball:2013hta}
Ball R~D, Bertone V, Carrazza S, Del~Debbio L, Forte S, Guffanti A, Hartland
  N~P and Rojo J (NNPDF) 2013 {\em Nucl. Phys. B\/} {\bf 877} 290--320
  (\textit{Preprint} \eprint{1308.0598})

\bibitem{Bertone:2013vaa}
Bertone V, Carrazza S and Rojo J (APFEL) 2014 {\em Comput. Phys. Commun.\/}
  {\bf 185} 1647--1668 (\textit{Preprint} \eprint{1310.1394})

\bibitem{Arbuzov:2006mu}
Arbuzov A~B and Scherbakova E~S 2006 {\em JETP Lett.\/} {\bf 83} 427--432
  (\textit{Preprint} \eprint{hep-ph/0602119})

\bibitem{Arbuzov:2019hcg}
Arbuzov A~B 2019 {\em Phys. Part. Nucl.\/} {\bf 50} 721--825

\bibitem{Ellis:1996nn}
Ellis R~K and Vogelsang W 1996  (\textit{Preprint} \eprint{hep-ph/9602356})

\bibitem{Gribov:1972rt}
Gribov V~N and Lipatov L~N 1972 {\em Sov. J. Nucl. Phys.\/} {\bf 15} 675--684

\bibitem{Curci:1980uw}
Curci G, Furmanski W and Petronzio R 1980 {\em Nucl. Phys. B\/} {\bf 175}
  27--92

\bibitem{Arbuzov:2002rp}
Arbuzov A 2003 {\em JHEP\/} {\bf 03} 063 (\textit{Preprint}
  \eprint{hep-ph/0206036})

\bibitem{Furmanski:1980cm}
Furmanski W and Petronzio R 1980 {\em Phys. Lett. B\/} {\bf 97} 437--442

\bibitem{Baikov:2012rr}
Baikov P~A, Chetyrkin K~G, Kuhn J~H and Sturm C 2013 {\em Nucl. Phys. B\/} {\bf
  867} 182--202 (\textit{Preprint} \eprint{1207.2199})

\bibitem{Gorishnii:1991hw}
Gorishnii S~G, Kataev A~L and Larin S~A 1991 {\em Phys. Lett. B\/} {\bf 273}
  141--144 [Erratum: Phys.Lett.B 275, 512 (1992), Erratum: Phys.Lett.B 341, 448
  (1995)]

\end{thebibliography}

\end{document}